\documentstyle[11pt,twoside]{article}
\begin{document}\hbadness=10000\thispagestyle{empty}
\pagestyle{myheadings}
\markboth{H.-Th. Elze, T. Kodama and J. Rafelski}
{The Sound of Sonoluminescence}
\title{{\bf The Sound of Sonoluminescence}}
\author{$\ $ \\
{\bf H.-Thomas Elze$^{1,2}$, Takeshi Kodama$^2$ and Johann
Rafelski$^1$}\\ $\ $\\
$^1$Department of Physics, University of Arizona, Tucson, AZ
85721\\ \rule{0cm}{0.7cm} and \\ \rule{0cm}{0.6cm} 
$^2$Universidade Federal do Rio de Janeiro, Instituto de
F\'{\i}sica\\
\,\,\,\,\,\,Caixa Postal 68.528, 21945-970 Rio de Janeiro, RJ, Brazil $\ $}
\date{August 10, 1997}
\maketitle 
\vspace{-8.5cm}
\vspace*{8.0cm}
\begin{abstract}{\noindent 
We consider an air bubble in water under conditions of single
bubble sonoluminescence (SBSL) and evaluate the emitted sound field
nonperturbatively for subsonic gas-liquid interface motion. Sound
emission being the dominant damping mechanism, we also implement
the nonperturbative sound damping in the Rayleigh-Plesset equation
for the interface motion. We evaluate numerically the sound pulse
emitted during bubble collapse and compare the nonperturbative and
perturbative results, showing that the usual perturbative
description leads to an overestimate of the maximal surface
velocity and maximal sound pressure. The radius vs. time
relation for a full SBSL cycle remains deceptively unaffected.\\
 
\noindent
43.28.R\,, 43.35.+d\,, 47.10.+g\,, 47.35.+i\,, 47.55.Bx\,, 47.55.Dz\,, 
68.10.-m\,
}\end{abstract}
\section{Introduction} 
This extensive study of the sound produced in single bubble cavitation
is prompted by the recent interest in sonoluminescence (SL), i.e.
the conversion of externally applied sound in a liquid into light
\cite{Bar97}. It has been generally accepted that the SL light
emission is intimately connected to the dynamics of the gas-liquid
interface, though the specific mechanism producing the light 
flashes has not been uniquely identified. We present here a
nonperturbative study of the sound radiated from the subsonically
moving phase boundary between the gas bubble and the liquid in
which it is immersed. In our approach we rely on the standard
treatise of the subject \cite{Pierce}. Apart from the importance of
sound emission as the primary damping mechanism under the extreme
conditions of sonoluminescence, the sound signal can be used as
another diagnostic tool to study the bubble dynamics. For detailed
accounts of related previous theoretical studies of the cavitation
sound see Refs.\,\cite{KellerMiksis,Loefstedtetal},
including references to earlier work. We proceed here
in two distinct steps: 
\begin{itemize} 
\item 
To begin with, the motion of the gas-liquid phase boundary
is assumed to be given and we determine the resulting sound
field.
\item 
Secondly, we consider the feedback effect of the radiated sound field
onto the phase boundary dynamics, i.e. we describe the motion of
the bubble selfconsistently. 
\end{itemize} 
 
The (multi-)bubble cavitation has been studied intensely for some time,
because of important applications and interesting underlying fluid
dynamics \cite{Neppiras,Young,Brennen}. The discovery of well
controlled experimental conditions by Gaitan and Crum \cite{Gai90}
permitting the study of a single gas bubble over long periods of
time, together with the ability to drive it externally by acoustic
waves, has focused the experimental interest on the single bubble
sonoluminescence (SBSL) phenomenon. The remarkable finding here is
that light is produced in very short pulses \cite{Bar91}: a million
photons of several eV energy are emitted within 50\,ps by
apparently about $10^{10}$ atoms/molecules from a gas bubble of
sub-$\mu$m radius. 
 
The standard theoretical tool in the study of the bubble
surface dynamics is the Rayleigh-Plesset (RP) equation, 
representing the motion of the gas-liquid interface by the 
time dependent (spherical) bubble radius $R(t)$\,. Several
mechanisms which damp the collective motion of the bubble-liquid
system have been incorporated in modern approaches: viscosity of
the liquid and sound damping \cite{Loefstedtetal}. Effects of the
`rectified' mass diffusion \cite{Bar97,Neppiras} and heat
conductivity, see e.g. Ref.\,\cite{ChuLeung}, play also an
important role in establishing the parameter regime for 
stationary SBSL, but will not be further discussed here. In
the SBSL cycle the sound emission produced during the
violent bubble collapse turns out to be the most important energy
loss mechanism \cite{Bar97,lightscatteringnew}. Less than half of
the energy in the bubble is dissipated by the viscous friction.
Only a comparatively tiny fraction of the energy delivered to the
bubble by the external sound wave is radiated in form of the
visible light \cite{Bar97,lightscatteringnew}. 
 
We will use the linear acoustic approximation (see Section 2.1),
which describes the subsonic propagation of density perturbations
in a fluid neglecting the effects of sound wave dispersion and
absorption \cite{Pierce}. This approximation does not allow us to
study situations involving shock wave formation in the liquid. In
Section 2.2 we solve the sound wave equation for the velocity
potential exactly for any given $R(t)$\,. Our derivation is
formally exact provided the assumptions of non-dissipative fluid
dynamics are valid. Our expressions account in full for retardation
and allow an arbitrary incoming sound field. The key result is the
exact form of the pressure amplitude radiated from the moving
bubble surface into the surrounding liquid, see
Eq.\,(\ref{pressure}). In order to compare with earlier analytical
and numerical work we also derive in Section 2.3 the perturbative
expansion of our general results in powers of $\dot R/c$, i.e.
bubble wall velocity divided by liquid sound velocity.
 
In the subsequent Section 3 we describe a model of the bubble
interior which is motivated by the approximately homologous
dynamics observed in numerical hydrodynamic simulations \cite{Chu}:
the gas density distribution and velocity field have been found to
show a simple scaling behavior, in terms of the scaled radius
variable $\xi\equiv r/R(t)$ and with $v(r,t)\approx\xi\dot R$,
respectively. These numerical results have recently motivated
development of a semianalytic approach for the coupled
bubble-liquid system, based on a Lagrangian variational principle
\cite{Sco97,Tak97}, of which a simpler variant will be used here.
This allows to assess the order of magnitude of the feedback
effects of the bubble interior dynamics.
 
With this preparation we can proceed to the second step indicated
above in Section 4, i.e. the derivation of a selfconsistent bubble
equation of motion. There exists already a variety of equations
generalizing the Rayleigh-Plesset equation, see e.g.
\cite{Neppiras,Young,Brennen}, and more recent ones which consider
the case of the externally driven damped nonlinear bubble
oscillations in the SBSL parameter regime
\cite{KellerMiksis,Loefstedtetal}. In our selfconsistent approach
(Section 4.1) we make extensive use of the results obtained in
Section 2\,. We incorporate the external sound field in Section 4.2
according to standard experimental conditions realized to date. In
Section 4.3 we derive a nonperturbative equation,
Eq.\,(\ref{Reqgen3}), which is formally valid to all orders in
$\dot R/c$ ($<1$), however, subject to the linear acoustic
approximation scheme. We incorporate the bubble interior using the
van\,der\,Waals equation of state and allow also for a realistic
equation of state (EoS) of the liquid. Our derivations are based on
the Navier-Stokes equation and, thus, take the effects of the
liquid and gas viscosities consistently into account.
 
We illustrate our formal results in Section 5 by studying
numerically the properties of an air bubble in water externally
driven by an ultrasound field with system parameters in the typical
SBSL regime \cite{Bar97,lightscatteringnew}. We show in detail
there that the various (nonperturbative) corrections incorporated
in the bubble equation of motion generally result in sizeable
corrections of the maximal bubble surface velocity during the
collapse. They tend to make the bubble collapse less violent than
seen in previous perturbative descriptions. In Section 5.3 we
compute the outgoing compression wave emitted from the bubble
surface into the surrounding liquid and consider the validity of
the acoustic approximation. We present final remarks and
conclusions accompanied by a brief summary of our work in Section
6\,.

\section{Sound Emission from Phase Boundaries}
We reexamine here the sound radiation originating from a moving
phase boundary between two non-viscous fluids, separated by an
idealized wall of vanishing thickness. References to earlier work
on this subject are surveyed in Refs.\,\cite{Pierce,Young,Brennen}.
Related issues were studied recently in
Refs.\,\cite{KellerMiksis,Loefstedtetal}. In distinction to these
approaches, we derive exact expressions for
the sound field emitted from a spherical boundary undergoing
arbitrary motion. We begin in the next subsection by recapitulating
some aspects of nonviscous hydrodynamics which lead to 
the (acoustic) approximation scheme used
here in describing the sound generation and propagation. 
\subsection{Acoustic Approximation}
The linear acoustic approximation is based on the 
assumption that the sound field
causes only small perturbations of the ambient state 
of the fluid characterized
by a vanishing velocity field and constant density and pressure Ref.\,\cite{Pierce}. 
Thus we shall not consider here nonlinear phenomena, such as the
dispersion and absorption of sound waves in the medium and,
in particular, the formation of supersonic shock waves. We shall see
quantitatively in Section 5 that this scheme is fairly well satisfied,
though the amplitude of the generated sound field typically reaches
up to $10^4$\,atm near to a SBSL bubble.
 
The basic equations of nonviscous hydrodynamics are:
\begin{enumerate}
\item The continuity equation for the fluid density $\rho\;$, which
assures the conservation of mass:
\begin{equation} \label{continuity}
 \partial_t\rho +\nabla\cdot (\rho\vec{v})=0 \;\;, 
\end{equation}
where $\vec{v}$ denotes the local fluid velocity.
\item The Euler equation, arising from Newton's equation of motion for
a small fluid cell:
\begin{equation} \label{Euler}
 \rho D_t\vec{v}=-\nabla P \;\;, 
\end{equation} 
where $P$ denotes the local pressure in the nonviscous fluid and
the comoving derivative is defined by 
$D_t\equiv\partial_t +\vec{v}\cdot\nabla$. Later we will consider
the Navier-Stokes generalization for viscous flow, see
Eq.\,(\ref{NavierStokes}).
\item In the absence of entropy production the set of equations is
closed by the EoS relating the pressure to the density. For example,
the ideal gas EoS is: 
\begin{equation} \label{EOS} 
 P[\rho]=K\rho^\gamma \;\;,\quad \gamma\equiv C_P/C_V \;\;,
\end{equation} 
where the adiabatic index $\gamma$ denotes the ratio of the specific
heat at constant pressure and volume, respectively.
\end{enumerate}
Had we allowed that the sound waves disturb the medium and generate
entropy, we would require an EoS with two variables, e.g relating
the pressure to the mass and energy densities. In this case we would
have to introduce additionally the energy conservation in differential
form, in order to close the set of equations.
 
In the following we study small-amplitude, acoustic perturbations
of an ambient state of the fluid (liquid, gas) characterized by the
ambient solutions to Eqs.\,(\ref{continuity})--(\ref{EOS}),
$P_0,\;\vec{v}_0,\;\rho_0$. In order to obtain equations governing
the perturbations $P^*,\;\vec{v}^{\;*},\;
\rho^*$, we set: 
\begin{equation} \label{perturb}
P=P_0+P^*\;\;,\quad \vec{v}=\vec{v}_0+\vec{v}^{\;*}\;\;,\quad 
\rho =\rho_0 +\rho^*\;\;. 
\end{equation} 
Inserting these expressions into 
Eqs.\,(\ref{continuity})--(\ref{EOS}), a coupled set of 
equations results which relates
various powers of the perturbations. The description of
perturbations can be further simplified by assuming a particularly
simple ambient state: 
\begin{equation} \label{ambient} 
P_0(\vec{x},t)=\mbox{Const.}\;\;,\quad
\vec{v}_0(\vec{x},t)=0\;\;,\quad
\rho_0(\vec{x},t)=\mbox{Const.}\;\;, 
\end{equation} 
i.e. a homogeneous state independent of time (quiescent state).
Then, the equations linearized in the perturbations become: 
\begin{eqnarray} \label{continuity1} 
&\;&\partial_t\rho^*+\rho_0\nabla\cdot\vec{v}^{\;*}=0
\;\;, \\ [2ex]
\label{Euler1} 
&\;&\rho_0\partial_t\vec{v}^{\;*}=-\nabla P^*
\;\;, \\ [2ex] 
\label{EOS1}
&\;&P^*=\gamma K\rho_0^{\gamma -1}\rho^*=c^2\rho^*\;\;,
\quad c^2\equiv\frac{\partial P}{\partial\rho}
\;\;,\end{eqnarray} 
where we made use of a Taylor expansion in order to obtain
Eq.\,(\ref{EOS1}) and introduced the (constant) sound velocity $c$
for the ambient fluid state. Combining 
Eqs.\,(\ref{continuity1})--(\ref{EOS1}), we obtain the well 
known wave equations: 
\begin{equation} \label{waves}
\partial_t^2P^*-c^2\Delta P^*=0\;\;,\quad 
\partial_t^2\rho^*-c^2\Delta\rho^*=0 
\;\;. \end{equation} 
 
Next, we introduce the velocity potential $\phi$\,. Consider the curl
of Eq.\,(\ref{Euler1}):
\begin{equation} \label{curlEuler}
\rho_0\partial_t\nabla\times\vec{v}^{\;*}=-\nabla\times\nabla P^*=0
\;\;, \end{equation} 
i.e. $\nabla\times\vec{v}^{\;*}(\vec x,t)=C(\vec x)$\,. 
Thus, if the vorticity of the velocity field perturbation $\vec
v^{\;*}$ vanishes initially, then it remains zero always. In
this case we can write:
\begin{equation} \label{potential} 
 \vec{v}^{\;*}\equiv\nabla\phi\;\;,\quad
 P^*\equiv -\rho_0\partial_t\phi 
\;\;, \end{equation}
where the relation between the velocity potential and the pressure
perturbation is chosen such that the linearized Euler equation,
Eq.\,(\ref{Euler1}), is automatically satisfied. Inserting the EoS, 
Eq.\,(\ref{EOS1}), in the linearized continuity equation,
Eq.\,(\ref{continuity1}), and using Eq.\,(\ref{potential}), one
obtains the wave equation for the velocity potential: 
\begin{equation} \label{potentialwave}
\partial_t^2\phi -c^2\Delta\phi =0 
\;\;. \end{equation} 
For simplicity, we suppress from now on the index $*$ introduced in
Eq.\,(\ref{perturb}). 

\subsection{Outgoing Spherical Wave Dynamics} 
Assuming spherical symmetry and using spherical coordinates, the
velocity potential is seen to satisfy the usual scalar spherical
wave equation:
\begin{equation} \label{sphwave} 
 0=r\{\partial_t^2\phi -\frac{c^2}{r}\partial_r^2(r\phi )\}
 =\partial_t^2(r\phi )-c^2\partial_r^2(r\phi ) 
\;\;. \end{equation} 
The generic solutions (regular at infinity) are out- and ingoing
spherical waves: 
\begin{equation} \label{sphwavesol} 
\phi_{out}=r^{-1}f(t-r/c)\;\;,\qquad
\phi_{in}=r^{-1}g(t+r/c) \;\;. \end{equation} 
For the spherically symmetric bubble wall located at the position
$r=R(t)$ the boundary condition is:
\begin{equation} \label{boundary} 
v(R,t)=\partial_r\phi (r,t)|_{r=R(t)}=\dot R(t)
\;\;, \end{equation}
where we introduced the notation $\dot X\equiv dX/dt$\,.
In general, the solution of the wave equation, 
Eq.\,(\ref{sphwavesol}), will be composed of a linear superposition
of in- and outgoing spherical waves. We shall allow for the
possibility of an incoming acoustical wave, $\phi_{in}$, e.g. as
part of the external driving sound field, and rewrite
Eq.\,(\ref{boundary}) in more detail: \begin{equation}
\label{boundary1} 
\partial_r\phi_{out}|_{r=R}=-\frac{1}{R^2}f(t-R/c)
 -\frac{1}{cR}f'(t-R/c)
=\dot R-\partial_r\phi_{in}|_{r=R} 
\equiv\dot {\cal R} \;\;, \end{equation} 
where $\phi_{out}$ has been substituted by an outgoing
spherical wave, cf. Eqs.\,(\ref{sphwavesol}); henceforth we 
use the abbreviation $f'(x)\equiv df/dx$\,. The external field
$\phi_{in}$ is kept arbitrary in the derivations to
follow, but will be incorporated according to a physically relevant
example in Section 4.2\,.
 
In principle, there will be also a sound field generated by the
moving phase boundary which will travel into the interior, to be
reflected at the center of the bubble ($r=0$), and then return. 
Its description, i.e. pertaining to the bubble interior, 
requires a suitable modification of the present
approach, see Section 3\,.
 
With $\dot {\cal R}(t)$ given by other dynamics, our  
objective now is to determine
implicitly the outgoing velocity potential $\phi_{out}$\,. For this
purpose we recast Eq.\,(\ref{boundary1}) into the form of an
ordinary linear first-order differential equation with known time
dependent coefficients:
\begin{equation} \label{boundary2} 
-\frac{1}{R^2}F-\frac{1}{cR}\frac{1}{1-\dot R/c}\dot F=
   \dot {\cal R} \;\;, \end{equation} 
where we used of the following substitutions and simple relations:
\begin{eqnarray} \label{tretard} 
&\;&t_-(t)\equiv t-R(t)/c\;\;\Longrightarrow\;\;
   \dot t_-=1-\dot R(t)/c \;\;, \\ [2ex] 
\label{fprime} 
&\;&f'(t-R/c)=\frac{df}{dt_-}\frac{dt_-}{dt} \frac{1}{\dot t_-}
=\frac{\dot f(t_-)}{1-\dot R/c} 
\;\;, \\ [2ex] \label{F} 
&\;&F(t)\equiv f(t_-(t))\;\;\Longrightarrow\;\;
    \dot F(t)=\dot f(t_-(t)) 
\;\;. \end{eqnarray} 
In Eq.\,(\ref{boundary2}) we note the retardation factor originating
in Eq.\,(\ref{fprime}). The general solution of Eq.\,(\ref{boundary2})
is elementary:
\begin{eqnarray} 
F(t)&=&F_0\frac{R(t)}{R(t_0)}\,\mbox{\large e}^{
  -\int_{t_0}^t{d}t'\frac{c}{R(t')}} 
 \nonumber \\ [2ex] \label{boundary2sol} 
&\;&-R(t)\int_{t_0}^t{d}t''[c-\dot R(t'')]\dot {\cal R}(t'')
\,\mbox{\large e}^{-\int_{t''}^t{d}t'\frac{c}{R(t')}}
\;\;, \end{eqnarray} 
where $F_0\equiv F(t_0)$ denotes the integration constant. It is
fixed by the requirement that no outgoing wave should be excited
before the bubble wall is set into motion at $t_0$\,. Therefore,
with $\dot {\cal R}(t<t_0)\equiv 0$, we find $F_0=0$\,. The resulting 
constraint $t>t_0$ is most
conveniently implemented by a step function factor $\theta(t-
t_0)$\,.
 
Using Eqs.\,(\ref{sphwavesol}),\,(\ref{F}), we can recover from
the above solution the outgoing velocity potential. To do this we
will need the inverse function $\tilde t(t_-)\equiv t$ of $t_-(t)$,
Eq.\,(\ref{tretard}). Note that $\tilde{t}$ is a single-valued
function, if and only if $t_-(t)=t-R(t)/c$ is either strictly
increasing or strictly decreasing with time. For subsonic bubble
wall motion $t_-(t)$ is strictly increasing, since $1-\dot
R/c>0$\,, and $\tilde{t}$ is strictly increasing in this case.
This allows us to perform variable substitution at will and we
obtain:
\begin{eqnarray}\label{phioutexp}
\phi_{out}(r,t)=\frac{1}{r}F(\tilde t(t-r/c))
  &=&-\theta(s-t_0)\frac{R(s)}{r}\int_{t_0}^{s}dt''
 [c-\dot R(t'')]\dot{\cal R}(t'')\,
\mbox{\large e}^{-\int_{t''}^{s}
       dt'\frac{c}{R(t')}}\;\;,\\ \label{defs}
s&\equiv&\tilde t(t-r/c)=t-\frac{r-R(s)}{c}\;\;.
\end{eqnarray}
The retardation effect seen here can be evaluated more explicitly. 
For example: 
\begin{eqnarray} \label{ttilde1} 
\tilde{t}(t-r/c)&\approx&t-\frac{r-R(t-r/c)}{c}
\;\;,\;\;\;r\gg R(t)\;\;, 
\\ [2ex] 
\label{ttilde2}  
\tilde{t}(t-r/c)&\approx&t-\frac{r-R(t)}{c} 
\;\;,\;\;\;r\stackrel{>}{\approx}R(t) 
\;\;. \end{eqnarray} 
 
In order to calculate the velocity and pressure fields, we require the
derivative of $\tilde t$: 
\begin{equation} \label{tretardinv} 
\tilde{t}(t-R(t)/c)\equiv t\;\;\Longrightarrow\;\;
\tilde{t}'(t_-)=\frac{1}{1-\dot R(\tilde{t}(t_-))/c} 
\;\;. \end{equation} 
Then, differentiating Eq.\,(\ref{phioutexp}), we obtain ($s>t_0$):
\begin{eqnarray} 
v(r,t)&=&\partial_r[\phi_{out}(r,t)+\phi_{in}(r,t)]
\nonumber \\ [2ex] \label{velocity} 
&=&\left\{\frac{R\dot {\cal R}}{r}+(\frac{1}{R}-
\frac{1}{r})\frac{F}{r} \right\}_{\tilde{t}(t-
r/c)}+\partial_r\phi_{in}(r,t) 
\;\;, \\ [4ex]
P(r,t)&=&-\rho_L\partial_t[\phi_{out}(r,t)+\phi_{in}(r,t)] 
\nonumber \\ [2ex] \label{pressure} 
&=&\frac{c\rho_L}{r}\left\{ R\dot {\cal R}
+\frac{F}{R}\right\}_{\tilde{t}(t-r/c)} 
-\rho_0\partial_t\phi_{in}(r,t) 
\;\;, \end{eqnarray}
where all functions within the braces are to be evaluated at the
indicated time-argument $\tilde t(t-r/c)=s$\,; henceforth $\rho_L$ 
denotes the
ambient density of the liquid. We recall that $P$ is the pressure
perturbation due to the generated sound field, to be added to the
ambient pressure. 

\subsection{Series Expansion for the Sound Field} 
The results obtained for the velocity potential and the velocity
and pressure fields, Eqs.\,(\ref{boundary2sol})--(\ref{pressure})
in Section 2.2\,, involve an integration over the history of the
bubble wall motion. Thus the corresponding nonlinear differential
equation describing it selfconsistently (Section 4) could be expected 
to become nonlocal in time (see, however, Section 4.3). 
In any case, for sufficiently slow
motion a local approximation can be justified, see
Refs.\,\cite{Loefstedtetal,Neppiras,Young,Brennen} and earlier
references therein; this approach leads to a popular generalization
of the RP equation. Therefore, our next objective is a systematic
expansion in powers of $\dot R/c$ of our general expressions, in
particular of the velocity potential, 
Eq.\,(\ref{phioutexp}). 
 
Let us consider an integral of the form (in our case $j=c/R\,,/,\,
h\propto(1-\dot R/c)\dot{\cal R}\,)$: 
\begin{equation} \label{integral} 
I(s)\equiv\int_{t_0}^s{d}t''h(t'')\exp\{-\int_{t''}^s{d}t'j(t')\}
\;\;, \end{equation} 
with $j>0$ and $s\ge t_0$\,, 
cf. Eq.\,(\ref{boundary2sol}). For a sufficiently large integrand
in the exponential the integration over $t''$ will be effectively
limited to a small range next to the upper limit $s$\,. This
suggests the following expansion: 
\begin{eqnarray} \label{integral2} 
I(s)&=&\int_{t_0}^s{d}t''h(t'')\exp\{-\int_{t''}^s{d}t'(j(t')-
j(s))\} \exp\{j(s)(t''-s)\}
\nonumber \\ [2ex] 
&=&\sum_{n=0}^\infty\frac{1}{n!}K^{(n)}(s)
\frac{{d}^n}{{d}j^n}\int_{t_0}^s{d}t''\exp\{j(s)(t''-s)\} \nonumber
\\ [2ex] 
&=&\sum_{n=0}^\infty\frac{1}{n!}K^{(n)}(s)
\frac{{d}^n}{{d}j^n}\frac{1-\exp\{j(s)(t_0-s)\}}{j(s)} \;\;,
\end{eqnarray} 
where: 
\begin{equation} \label{h} 
K^{(n)}(s)\equiv\left.\frac{{d}^n}{{d}\hat{t}^n}
 h(\hat{t})\exp\{-\int_{\hat{t}}^s{d}t'(j(t')-j(s))\}\right
|_{\hat{t}=s} \;\;. \end{equation} 
We remark that the exponentially small correction term (for $s\gg
t_0$) on the r.h.s. of Eq.\,(\ref{integral2}) can be attributed to
transient contributions to the integral or, rather, to the function
$F$, Eq.\,(\ref{boundary2sol}), which are due to the switching-on
of the bubble wall motion considered in Section 2.2\,. Since we
will be mainly interested in the nonlinear oscillatory motion of
the bubble wall when all transients have died out, we can neglect
this term for sufficiently late times. 
Thus, we obtain: \begin{equation} \label{integral3} 
I(s)=\sum_{n=0}^\infty (-1)^nK^{(n)}(s)j^{-n-1}(s) 
\;\;, \end{equation} 
which presents the starting point of our approximate evaluation of
the velocity potential. 
 
Employing Eqs.\,(\ref{boundary2sol}),\,(\ref{phioutexp}) from the
previous section together with
Eqs.\,(\ref{integral}),\,(\ref{integral3}) above, we obtain: 
\begin{eqnarray} \label{phiapprox} 
&\phantom{.}&\phi_{out}(r,t)\,= 
\\ [2ex] 
&\phantom{.}&-\frac{R^2(s)}{r}\sum_{n=0}^\infty 
 (-\frac{R(s)}{c} )^n\left . 
\frac{{d}^n}{{d}\hat{t}^n} (1-\frac{\dot R(\hat{t})}{c} ) \dot
{\cal R}(\hat{t})\exp\{(s-\hat{t})\frac{c}{R(s)}+
\int_{s}^{\hat{t}}{d}t\frac{c}{R(t)}\}\right |_{\hat{t}=s} 
\nonumber \\ [2ex] 
&=&-\frac{R^2}{r}\dot {\cal R}\left\{ 
1-2\frac{\dot R}{c}(1+\frac{1}{2}\frac{R\ddot {\cal R}}{\dot R\dot
{\cal R}}) +2(\frac{\dot R}{c})^2(1+2\frac{R\ddot {\cal R}}{\dot
R\dot {\cal R}}+ \frac{R\ddot R}{\dot R^2}+\frac{1}{2}\frac{R^2
\stackrel{...}{\cal R}
}{\dot R^2 \dot {\cal R}})+{\rm O}\left ((\dot R/c)^3\right
)\right\}_{s} \;, \nonumber \end{eqnarray} 
where as before $s=\tilde{t}(t-r/c)$\, see Eq.\,(\ref{defs}). The
expansion has to be carried out up to order $n=4$ in order to
acquire in the $\dot R/c$ expansion all second order terms. Our result
reduces to Eq.\,(18) of Ref.\,\cite{Loefstedtetal} at the ${\rm
O}(\dot R/c)$ obtained there, provided we replace i) ${\cal
R}(t)\longrightarrow R(t)$, i.e. we ignore effects of an incoming
sound field, cf. Eq.\,(\ref{boundary1}), and ii) $\tilde{t}(t-
r/c)\longrightarrow t-[r-R(t)]/c$\,. The latter approximation
amounts to neglecting already some ${\rm O}(\dot R/c)$ corrections
in Eq.\,(\ref{phiapprox}) due to retardation effects.
 
To conclude this subsection, we calculate the velocity and pressure
fields generated by a driven oscillating bubble wall, i.e. the
sound emitted, by combining  Eqs.\,(\ref{phioutexp}),
(\ref{velocity}) and (\ref{pressure}), respectively, and
Eq.\,(\ref{phiapprox}): 
\begin{eqnarray} \label{velocityout} 
&\phantom{.}&v(r,t)\;=\;\partial_r\phi_{in}(r,t) 
+\frac{R^2\dot {\cal R}}{r^2}
\\ [2ex] 
&\phantom{.}&\;\;+\;2\frac{\dot R}{c}\frac{R\dot {\cal R}}{r} \left
(1-\frac{R}{r}\right ) 
\left\{
1+\frac{1}{2}\frac{R\ddot {\cal R}}{\dot R\dot {\cal R}}
-\frac{\dot R}{c}(1+2\frac{R\ddot {\cal R}}{\dot R\dot {\cal R}}+
\frac{R\ddot R}{\dot R^2}+\frac{1}{2}\frac{R^2
\stackrel{...}{\cal R}
}{\dot R^2
\dot {\cal R}})+{\rm O}\left ((\dot R/c)^2\right )\right\}
_{\tilde{t}(t-r/c)} 
, \nonumber \end{eqnarray} 
which illustrates nicely that our expansion reproduces the
incompressible fluid ($c\rightarrow\infty$) limit; furthermore: 
\begin{eqnarray} \label{pressureout} 
P(r,t)&=&-\rho_L\partial_t\phi_{in}(r,t) 
\\ [2ex] 
&+&\frac{2\rho_L}{r}R\dot R\dot {\cal R}
\left\{
1+\frac{1}{2}\frac{R\ddot {\cal R}}{\dot R\dot {\cal R}}
-\frac{\dot R}{c}(1+2\frac{R\ddot {\cal R}}{\dot R\dot {\cal R}}+
\frac{R\ddot R}{\dot R^2}+\frac{1}{2}\frac{R^2
\stackrel{...}{\cal R}
}{\dot R^2
\dot {\cal R}})+{\rm O}\left ((\dot R/c)^2\right )\right\}
_{\tilde{t}(t-r/c)} 
\;, \nonumber \end{eqnarray} 
where the inverse of the retarded time function has to be inserted
everywhere as indicated before.

\section{Homologous Bubble Interior Dynamics}
In section 4 we want to derive the equation of motion of the phase
boundary (bubble surface) self-consistently. For this purpose we
need to understand the dynamics of the interior of the bubble. In
distinction to Section 2\,, presently we have to take into account
that the density, pressure, and velocity fields inside the bubble
may change by several orders of magnitude during different phases
of the (periodic) bubble motion. Therefore, a linearization of the
hydrodynamic equations of motion around a homogeneous and time
independent ambient state, as performed in Section 2.1\,, is not
applicable here. 
 
In the hydrodynamic studies of the interior motion the emphasis has
been to understand the development of the extreme conditions inside
the bubble \cite{WuR93,Mos94}. Chu noted that for an important part  
of the SBSL cycle the bubble motion is (nearly) homologous, i.e.
the shape of the density distribution and the velocity field of the
gas scale in an appropriate way with the radius $R$ and surface
velocity $\dot{R}$ \cite{Chu}. The numerical simulations in the
SBSL parameter regime indicate a homologous contraction (and its
stability) until the onset of shock wave formation and
independently of the details of the EoS used. This stability of the
homologous motion may be related to the fact that only for the
homologous motion there is no energy loss due to viscosity inside
the bubble (see Eq.\,(\ref{viscinside}) and below).
 
We solve the continuity and Euler equations,
Eqs.\,(\ref{continuity}) and (\ref{Euler}) of Section 2.1
respectively, by approximating the density and velocity fields with
suitable scaling functions. We neglect dissipative effects in the
bubble, such as viscosity, heat conduction, and radiation transfer
especially. Presently, we are interested in a semiquantitative
analysis of the bubble interior. This seems sufficient for our
later derivation of the overall bubble dynamics: the motion is
largely determined by the dynamics of the fluid, since only a minor
fraction of the relevant total energy resides in the interior at
any time. For a detailed understanding of the microscopic processes
leading to sonoluminescence inside the bubble, however, the
dissipative effects will be essential ingredients. 
 
It turns out, see Refs.\,\cite{Sco97,Tak97}, that for a time independent 
shape of a properly scaling density distribution the velocity 
potential is:
\begin{equation} \label{velpot} 
\phi (r,t)\equiv \frac{1}{2}\frac{\dot R}{R}r^2\;\;,
\quad r\le R\;\;, \end{equation} 
in terms of the bubble radius $R(t)$, which yields: 
\begin{equation} \label{dvelpot} 
v=\partial_r\phi =\frac{\dot R}{R}r\;\;,\quad
\partial_t\phi =\frac{1}{2}
\left(\frac{\ddot R}{R}-(\frac{\dot R}{R})^2 \right )r^2
\;\;, \end{equation} 
i.e. a linear radial velocity profile. The continuity equation,
Eq.\,(\ref{continuity}), takes the form:
\begin{equation} \label{conteq} 
0=\partial_t\rho +\frac{1}{r^2}\partial_r(r^2\rho v) 
=\left (\dot\varrho +3\frac{\dot R}{R}\varrho\right )\tilde{d}
+\varrho\partial_t\tilde{d}|_\xi 
\;\;, \end{equation} 
where we introduced the homologous ansatz:
\begin{equation} \label{densans} 
\rho (r,t)\equiv\varrho (t)\tilde{d}(\xi ,t) 
\;\;, \end{equation} 
with $\xi\equiv r/R(t)$ denoting the scaling variable. 
Here we find indeed that the velocity potential, Eq.\,(\ref{velpot}), 
is consistent with a 
time independent density profile function, i.e.
\begin{equation} \label{densshape} 
\tilde{d}(\xi ,t)=d(\xi )\;\;,
\end{equation} 
and the overall density factor of Eq.\,(\ref{densans}),
normalized to $N$ particles inside the bubble: 
\begin{equation} \label{denssol} 
\varrho (t)= \frac{N}{4\pi R^3(t)\int_0^1{d}\xi\;\xi ^2d(\xi )}
\;\;. \end{equation} 
We choose $d(1)\equiv 1$ in the following, i.e. $\rho
(R,t)=\varrho (t)$\,. We refer the reader to
Refs.\,\cite{Sco97,Tak97} for the variational
improvement of the strictly homologous dynamics (see \cite{Chu} and
references therein) which we will pursue here, with a time independent
yet scaling density profile function. 
 
In order to determine the scaling function $d(\xi)$, we consider
the integral of the Euler equation, cf. Eq.\,(\ref{Euler}), which for
our radially symmetric case yields:
\begin{equation} \label{Eulerint} 
\partial_t\phi |_R^r+\frac{1}{2}(\partial_r\phi )^2|_R^r
  =-\int_R^r{d}r \;\frac{1}{\rho}\partial_rP
\;\;. \end{equation} 
In the absence of entropy $S$ production the integral on the r.h.s. 
is the enthalpy:
\begin{equation} \label{enth1}
\int_r^R{d}r \frac{1}{\rho}\partial_rP
=\int_r^R \frac{dP}{\rho}\vert_S=h(R)-h(r)\,,\quad h={\cal E}
+\frac{P}{\rho}
\;\;, \end{equation}
where ${\cal E}={E}/{N}$ is the specific energy per particle.
 
During the highest compression particle
densities reached in the bubble interior 
are similar to those in the liquid outside. We thus
use here for the gas an adiabatic EOS, cf. 
Eq.\,(\ref{EOS}), with van\,der\,Waals hard core corrections: 
\begin{equation} \label{EOSW} 
P(\rho )=P_0\left (\frac{\rho}{\rho_0}\right )^\gamma
\left (\frac{\rho_a-\rho_0}{\rho_a-\rho}\right )^\gamma
\equiv\tilde{P}_0 \left (
\frac{\rho /\rho_0}{1-\rho /\rho_a}\right )^\gamma 
\;\;. \end{equation} 
Here $\rho_0\;(\ll\rho_a)$ is the density at which the bubble
interior is at the ambient pressure $P_0$ and $\rho_a^{-1}
\equiv\frac{4\pi}{3}a^3/N$ in terms of the van\,der\,Waals excluded
volume. The adiabatic index $\gamma$ for ideal monatomic (diatomic)
gases is 5/3 (7/5)\,. Related EoS have been applied to calculate
the gas pressure before \cite{Loefstedtetal,Sco97,WuR93}.
 
Performing the integral in Eq.\,(\ref{Eulerint}) using
Eq.\,(\ref{EOSW}) and employing the scaling form for the velocity
potential, Eq.\,(\ref{velpot}), we obtain ($\xi\equiv r/R(t)$): 
\begin{eqnarray} \label{Eulerint1}
\frac{1}{2}R\ddot R(\xi ^2-1)&=&
-\frac{\tilde{P}_0}{\rho_0} 
\left (\frac{\rho /\rho_0}{1-\rho /\rho_a}\right )^{\gamma -1}
\left.\left (\frac{1}{1-\rho /\rho_a}+\frac{1}{\gamma -1}\right )
\right |_R^r\;\;, 
\\ [2ex] 
\label{lowdens} 
&=&-\frac{\tilde{P}_0}{\rho_0}\left (\frac{\rho}{\rho_0}\right
)^{\gamma -1} \left.\left (\frac{\gamma}{\gamma -1}+{\rm O}(\rho
/\rho_a)\right ) \right |_R^r\;\;, 
\\ [2ex] 
\label{highdens} 
&=&-\frac{\tilde{P}_0}{\rho_a}\left.\left
(\frac{\rho_a}{\rho_0}\right )^\gamma 
\frac{1}{(1-\rho /\rho_a)^\gamma}
\left (1+{\rm O}(1-\rho /\rho_a) \right )\right |_R^r
\;\;, \end{eqnarray} 
where Eq.\,(\ref{lowdens}) and Eq.\,(\ref{highdens}), respectively,
represent the low- and high-density limits of the r.h.s. of
Eq.\,(\ref{Eulerint1}). 
  
Obviously, the van\,der\,Waals correction introduces an additional
scale which makes it difficult to find a universal density shape
function. Therefore, we consider the low- and high-density cases in
turn:
\begin{itemize} 
\item {\bf Low density.} 
Inserting Eqs.\,(\ref{densans})--(\ref{denssol}) into
Eq.\,(\ref{lowdens}), we obtain:
\begin{equation} \label{Eulerlow} 
\frac{\gamma -1}{2\gamma}\frac{\rho_0R\ddot R}{\tilde{P}_0}\left (
\frac{\rho_0}{\varrho (t)}\right )^{\gamma -1}=q_{\rm l}=
\frac{d^{\gamma -1}(\xi )-1}{1-\xi ^2}
\;\;. \end{equation} 
Since the l.h.s. here only depends on $t$ and the r.h.s. only on $\xi$,
we have introduced a constant parameter $q_{\rm l}$ characterizing
the time independent density profile.
Solving the r.h.s., we find:
\begin{equation} \label{flow} 
d_{\rm l}(\xi )=\left (1+q_{\rm l}(1-\xi^2)\right )
^{\frac{1}{\gamma -1}}
\;\;. \end{equation} 
We note that the l.h.s. of Eq.\,(\ref{Eulerlow}) determines 
the value of the parameter $q_{\rm l}$, and thus 
we can verify if indeed it
is time independent. From this we conclude that this low-density
($\rho\ll\rho_a$) functional form of the density profile function is
approximately valid for inertial motion with $\ddot R\approx 0$ or,
more generally, when the bubble wall accelerates (decelerates)
proportional to the internal pressure acting on it,
$P(R,t)\approx\tilde{P}_0(\varrho (t)/\rho_0)^\gamma$, that is
whenever the ratio $\{\varrho R\ddot R\} (t)/P(R,t)$ is
approximately constant.
\item {\bf High density.} 
In this limit, inserting Eqs.\,(\ref{densans})--(\ref{denssol})
into Eq.\,(\ref{highdens}), we obtain:
\begin{equation} \label{Eulerhigh} 
\frac{1}{2\gamma}\left ( 
\frac{\rho_0}{\rho _a}\right )^\gamma 
\frac{\rho_aR\ddot R}{\tilde{P}_0}
\frac{(1-\varrho (t)/\rho_a)^{\gamma +1}}{\varrho
(t)/\rho_a}=q_{\rm h}= \frac{d(\xi )-1}{1-\xi^2} 
\;\;, \end{equation} 
where again the l.h.s. is only a function of time, while the r.h.s. is
only a function of $\xi$, and thus we can introduce the constant
$q_{\rm h}$. In deriving Eq.\,(\ref{Eulerhigh}), we employed an
additional expansion for $d(\xi )\approx 1$ on the r.h.s. which is
consistent with the high-density case ($\rho\approx\rho_a$) under
consideration. Solving the r.h.s. for the density profile function, we
obtain: 
\begin{equation} \label{fhigh} 
d_{\rm h}(\xi )=1+q_{\rm h}(1-\xi^2)
\;\;,\end{equation} 
while the l.h.s. determines the value of the parameter $q_{\rm h}$, as 
before. The limit of validity of this result is controlled by the
requirement that the expression on the l.h.s. remains time
independent. Note that the high- and low-density profiles agree with 
each other for $\gamma=2$; however, $\gamma= 3/2$ and 5/2 for
mono- and diatomic gases, respectively.
\end{itemize}
We remark that the numerical values of $q_{\rm l}$ and 
$q_{\rm h}$, in general, vary depending on the dynamical 
regime, and always $q\ge -1$. It is important to realize that
Eqs.\,(\ref{Eulerlow}) and (\ref{Eulerhigh}) present implicit
equations for $q_{\rm l}$ and $q_{\rm h}$, respectively. The reason
for this is that $\varrho (t)$ because of the normalization in
Eq.\,(\ref{denssol}) depends on the integrated density profile and
thus on the respective $q$ by Eqs.\,(\ref{flow}),\,(\ref{fhigh}).
In practice, these `constants' will be treated as adiabatically
changing parameters to be computed selfconsistently from the
equations derived here. 
 
We observe that as the bubble goes through the cycle determined by
the driving pressure, the shape of the matter distribution inside
the bubble changes, controlled by $q$, which remains nearly
constant during much of the cycle and changes rapidly in
sign near to the bubble collapse/bounce, whenever $\ddot R=0$\,,
where the l.h.s. of 
Eqs.\,(\ref{Eulerint1}),\,(\ref{Eulerlow}), or (\ref{Eulerhigh})
vanishes, and $q=0$, assuring in this transition instant a (nearly)
homogeneous density distribution. For negative values of $q$ the
highest density is found at the surface, corresponding to a
collapsing phase; similarly, positive values of $q$ with the
highest density at the center correspond to an expanding phase of
the bubble motion. We found that there is an approximate fixed point
$r/R\equiv\xi _*$ in the motion, where the average density
corresponding to a homogeneous bubble is maintained during 
the entire cycle: 
\begin{equation}\label{rhostar}
\rho_*\equiv\rho(\xi_*R,t)\simeq N/(4\pi/3\, R(t)^3\;\;, 
    \quad \xi_*\approx 0.78\;\;.
\end{equation}
This will turn out to be a very useful property which we shall
exploit deriving specific numerical results in Section 5.2\,, where
we compare the bubble dynamics for a homogeneous and a homologous
interior, respectively. 

\section{Derivation of the Bubble Equation of Motion}
Having studied the effects of a given bubble wall motion on the
exterior liquid and the bubble interior in Sections 2 and 3\,,
respectively, we now turn to the question how this motion can be
determined selfconsistently. That is, assuming an external driving
sound field $\phi_{in}$, we want to derive an equation of motion
for the bubble radius $R$ as a function of time. We also take the
bubble interior into account. Its dynamics may have little
influence on most of the cycle of stable bubble oscillations.
However, in order to improve the understanding of the final stage
of the violent collapse \cite{Sco97,Tak97}, it has to be
considered. Besides the sound emission from the moving phase
boundary, which constitutes a major energy loss mechanism, the
effect of viscous damping will also be incorporated here, which is
of a similar order of magnitude as acoustic damping. Despite the
fact that our considerations so far were based on the assumption of
nonviscous fluid dynamics, i.e. approximately free wave
propagation in particular, we can include dissipative and driving
forces into the equation of motion for $R(t)$ by deriving it from
the Navier-Stokes (momentum balance) equation. Thus, only 
the backreaction of viscosity (and other dissipative transport
effects) on the wave propagation and velocity potential $\phi$ is
neglected here.

\subsection{The (Generalized) Rayleigh-Plesset Equation}
We begin with the Navier-Stokes equation generalizing the Euler
equation, see Eq.\,(\ref{Euler}), to the case of viscous fluids
\cite{Pierce}. For our spherically symmetric situation this
equation can be written in the form: 
\begin{equation} \label{NavierStokes} 
\partial_r\left\{\partial_t\phi +\frac{1}{2}(\partial_r\phi
)^2\right\} =-\frac{1}{\rho}\partial_rP+
\frac{\eta}{\rho}\partial_r\left (\frac{1}{r}\partial_r^2(r\phi
)\right ) \;\;, \end{equation} 
where $\phi (r,t)$ denotes the appropriate velocity potential, i.e.
for the interior or exterior region, and
$\eta\equiv\frac{4}{3}\eta_s+\eta_b$ is the relevant combination of
{\it s}hear and {\it b}ulk viscosities of the fluid entering here.
Similarly, the pressure $P$ has to be specified differently
according to whether the interior or exterior of the bubble is
considered. 
 
Due to the phase change at the bubble surface, there arises a
nontrivial boundary condition relating the normal components of the
stress tensor $\Sigma$ inside ($G$as) and outside ($L$iquid):
\begin{equation} \label{stressboundary}
\vec{n}\cdot \Sigma_G\cdot\vec{n}=
\vec{n}\cdot \Sigma_L\cdot\vec{n}-\frac{2\sigma}{R} 
\;\;, \end{equation} 
where the term $\propto\sigma$ denotes the `pressure' contribution
due to the (liquid) surface tension for the liquid/gas interface
and the viscous stress tensor has to be evaluated for the
respective fluid under consideration \cite{Pierce}: 
\begin{equation} \label{stresstensor} 
\vec{n}\cdot \Sigma\cdot\vec{n}\equiv\left\{
-P+\eta\partial_rv-\frac{\bar{\eta}}{r}
v\right\}_{r=R} 
\;\;, \end{equation} 
where $\bar{\eta}\equiv\frac{4}{3}\eta_s-2\eta_b$, $\eta$ was
defined after Eq.\,(\ref{NavierStokes}), and $v(r,t)\equiv
\partial_r\phi(r,t)$, as usual. Then, employing the boundary
condition (\ref{boundary}) and the velocity potential
(\ref{velpot}) for the gas phase, we obtain instead of
Eq.\,(\ref{stressboundary}) more explicitly: 
\begin{equation} \label{stressboundary1} 
-P_G(R)+(\eta_G-\bar{\eta}_G)\frac{\dot R}{R} 
=-P_L(R)+\eta_L\partial_rv|_{r=R}-\bar{\eta}_L\frac{\dot R}{R} -
\frac{2\sigma}{R} 
\;\;. \end{equation} 
Using Eq.\,(\ref{tretardinv}) and the velocity field 
(\ref{velocityout}), we 
calculate $\partial_rv|_R$ for the outer liquid: 
\begin{eqnarray} \label{dvelocitydr} 
\partial_rv(r,t)|_{r=R}&=&-2\frac{\dot R}{R} 
-\dot R^{-1}\left\{ 
[-2\frac{\dot R}{R}+\frac{\dot R}{c}\partial_t+ 
(\frac{\dot R}{c})^2\partial_t]\; 
\partial_r\phi_{in}(r,t)\right\}_{r=R} 
\nonumber 
\\ [2ex] 
&\;&-\; 2\frac{\dot {\cal R}}{R}
(\frac{\dot R}{c})^2
\left (1+2\frac{R\ddot {\cal R}}{\dot R\dot {\cal R}}+ 
\frac{R\ddot R}{\dot R^2}+\frac{1}{2}\frac{R^2
\stackrel{...}{\cal R}
}{\dot R^2
\dot {\cal R}}\right )+{\rm O}\left ((\dot R/c)^3\right )
\;\;, \end{eqnarray} 
where the first term on the RHS corresponds to the incompressible
fluid limit ($c\rightarrow\infty$) in the absence of an external
driving sound field.

Several remarks are in order here: 
\begin{itemize} 
\item 
The bubble gas pressure at the surface, $P_G(R)\equiv
P_G(\rho_G(R,t))$, is determined by the van\,der\,Waals EoS
Eq.\,(\ref{EOSW}) together with $\rho_G(R,t)=\varrho_G(t)$,
according to Eqs.\,(\ref{densans}),\,(\ref{denssol}) for the
homologous density function. The liquid vapor pressure inside the
bubble is not considered at present.
\item 
We observe that $\eta_G-\bar{\eta}_G =3\eta_{bG}$, in terms of the
bulk viscosity of the gas, which is generally very small compared
to liquid viscosities. The shear viscosity does not contribute here
at all, cf. also the remark after Eq.\,(\ref{viscinside}) below.
For completeness, we allow for viscous damping in the gas inside
the bubble, even though it might at best become important during
the rapid collapse and high compression phase.
\item
We recall that by definition of the inverse of the retarded time
function, Eq.\,(\ref{tretardinv}), we have $\tilde{t}(t-
R(t)/c)\equiv t$\,; therefore, the r.h.s. of
Eq.\,(\ref{dvelocitydr}) has to be evaluated at the time $t$\,.
\end{itemize} 
 
In the following the boundary condition (\ref{stressboundary1})
together with Eq.\,(\ref{dvelocitydr}) will be employed to
eliminate $P_L(R)$ in terms of the other quantities which by now
are explicitly calculated functions of $R,\;\dot R,\;\ddot R$ etc.
In order to proceed, we convert the Navier-Stokes equation
(\ref{NavierStokes}) together with the boundary condition
(\ref{stressboundary1}) into an ordinary differential equation for
$R(t)$. This can be achieved by integrating over the radial
coordinate, i.e. the approach advocated earlier in
Ref.\,\cite{KellerMiksis}, for example. Symbolically, in obvious
correspondence with the terms on the left- and right-hand sides of
Eq.\,(\ref{NavierStokes}), we obtain: 
\begin{equation} \label{symbol} 
\{\int_0^R+\int_R^\infty\}{d}r\;{\cal D}= 
\{\int_0^R+\int_R^\infty\}{d}r\;{\cal P}+ 
\{\int_0^R+\int_R^\infty\}{d}r\;{\cal V} 
\;\;, \end{equation} 
where we split the integrations at $r=R$ because of the different
phases inside and outside. Making use of the results of the
previous Sections 2 and 3, we proceed to evaluate each term of the
generalized Bernoulli equation (\ref{symbol}) in turn. 
 
Using the velocity potential (\ref{velpot}) for the bubble
interior, we obtain: 
\begin{eqnarray} \label{kininside} 
\int_0^R{d}r\;{\cal D}&\equiv& 
\int_0^R{d}r\; 
\partial_r\left\{\partial_t\phi +\frac{1}{2}(\partial_r\phi
)^2\right\} =\frac{1}{2}R\ddot R 
\;\;, \\ [2ex] 
\label{viscinside} 
\int_0^R{d}r\;{\cal V}&\equiv& 
\int_0^R{d}r\; 
\frac{\eta_G}{\rho}\partial_r\left (\frac{1}{r}\partial_r^2(r\phi
)\right ) =0 
\;\;. \end{eqnarray} 
We observe that the homologous scaling solution for the bubble
interior, $\phi\propto r^2$, does not give a contribution to the
integrated viscous force term here. This remarkable result suggests
that viscous forces drive the bubble motion to the homologous limit
which we discussed in Section 3 and which is in this respect quite
unique \cite{Chu,Sco97,Tak97}.
 
Furthermore, employing the van\,der\,Waals EoS (\ref{EOSW}), we
calculated the enthalpy integral for the gas inside the bubble
before, cf. Eqs.\,(\ref{enth1}),\,(\ref{Eulerint1}):
\begin{equation} \label{enthinside} 
\int_0^R{d}r\;{\cal P}\equiv 
-\int_0^R{d}r 
\;\frac{1}{\rho}\partial_rP_G= 
-\left . P_G(\rho_G)\frac{\gamma -\rho_G/\rho_a}
{(\gamma -1)\rho_G}\right |_0^R\equiv h_G[P_G] 
\;\;, \end{equation}
where we rewrote our previous result of Eq.\,(\ref{Eulerint1}) in
terms of the van\,der\,Waals expression for the pressure and denote
by $h_G$ the gas enthalpy.
 
Next, we turn to the evaluation of the corresponding integrals for
the exterior region. The first integral follows immediately from
our derivations in Section 2\,:
\begin{equation} \label{kinoutside} 
\int_R^\infty{d}r\;{\cal D}\equiv 
\int_R^\infty{d}r\; 
\partial_r\left\{\partial_t\phi +\frac{1}{2}(\partial_r\phi
)^2\right\} =\frac{1}{\rho_L}P(R,t)-\frac{1}{2}\dot R^2
\;\;, \end{equation} 
where $P(r,t)$ denotes the pressure field calculated in
Eq.\,(\ref{pressure}) and evaluated approximately in
Eq.\,(\ref{pressureout}); henceforth we denote by $\rho_L$ the
ambient density of the liquid. The second term on the r.h.s. of
Eq.\,(\ref{kinoutside}) simply follows from the boundary condition
(\ref{boundary}) and the assumption $\phi\rightarrow 0$
(sufficiently fast) for $r\rightarrow\infty$\,. 
 
In order to calculate the enthalpy integral for the liquid, we
employ an EoS which gives a realistic description for many liquids
\cite{Neppiras}: 
\begin{equation} \label{EOSL} 
P_L(\rho )=(P_0+P_1)\left (\frac{\rho}{\rho_L}\right )^n-P_1\;\;,
\;\;\;\frac{\partial P_L}
{\partial\rho}=c^2(\rho )
\;\;, \end{equation} 
where $P_0$ denotes the ambient pressure and $n$,\,$P_1$ are
parameters depending on the liquid (e.g. for water
$n=7$,\,$P_1=3$\,kbar). Then, it is straightforward to obtain:
\begin{equation} \label{enthoutside} 
\int_R^\infty{d}r\;{\cal P}\equiv 
-\int_R^\infty{d}r 
\;\frac{1}{\rho}\partial_rP_L= 
\frac{n}{n-1}\frac{P_0+P_1}{\rho_L}\left\{
\left (\frac{P_L(R)+P_1}{P_0+P_1}\right )^{\frac{n-1}{n}}-1\right\}
\equiv h_L[P_L] 
\;\;, \end{equation} 
using $P(r\rightarrow\infty)=P_0$\,, i.e. the ambient pressure, and
where $h_L$ denotes the liquid enthalpy. As mentioned before, the
liquid pressure on the bubble surface, $P_L(R)$, will be eliminated
via the pressure boundary condition (\ref{stressboundary1}). 
 
The last (viscous) term from Eq.\,(\ref{symbol}) can be evaluated
employing the approximation $\rho\approx\rho_L$:
\begin{equation} \label{viscoutside} 
\int_R^\infty{d}r\;{\cal V}\;\equiv\; 
\int_R^\infty{d}r\; 
\frac{\eta_L}{\rho}\partial_r\left (\frac{1}{r}\partial_r^2(r\phi
)\right ) =-\frac{\eta_L}{\rho_L}\frac{1}{r}\partial_r^2(r\phi
)|_{r=R}  =-\frac{\eta_L}{\rho_L}\left\{2\frac{\dot
R}{R}+\partial_rv|_{r=R} \right\}
\;\;, \end{equation} 
i.e. neglecting a cross term between viscosity and compressibility
of the liquid, which will be justified by the numerical results
presented in Section 5.3\,.
 
It is worth while to recall that in
Eqs.\,(\ref{kininside})--(\ref{viscoutside}) all the functions
$R,\,\dot R,\,\ddot R$, etc., which appear on the r.h.s.\,, are to
be evaluated simply at the time $t$\,.
 
Collecting the essential results from Eqs.\,(\ref{kininside})--
(\ref{viscoutside}), we obtain the equation of motion for the
bubble radius $R(t)$\,, i.e. $\int_0^\infty {d}r{\cal D}=
\int_0^\infty {d}r({\cal P+V})$\,, in the form: 
\begin{equation} \label{Req} 
\frac{1}{2}R\ddot R 
-\frac{1}{2}\dot R^2
+\frac{1}{\rho_L}P(R,t)=h_G[P_G]+h_L[P_L]  
-\frac{\eta_L}{\rho_L}\left\{2\frac{\dot R}{R}+\partial_rv|_{r=R}
\right\} 
\;\;. \end{equation} 
This equation is valid for any subsonic bubble wall velocity, it is
nonperturbative in $\dot R/c$\,. In fact, inserting here our
general results of Section 2.2 for the sound field generated by the
bubble, in particular $v(r,t)$ and $P(r,t)$ calculated in
Eqs.\,(\ref{velocity}) and (\ref{pressure}) respectively, we obtain
an important generalization of the Rayleigh-Plesset equation.
 
In order to illustrate the contents of Eq.\,(\ref{Req}), we proceed
with a perturbative evaluation of terms involving the generated
sound field making use of our results of Section 2.3\,. Using in
Eq.\,(\ref{Req}) the expansions in $\dot R/c$, especially
Eqs.\,(\ref{pressureout}),\,(\ref{dvelocitydr}), 
we obtain the perturbative bubble equation of motion:
\begin{eqnarray} \label{Reqperturb} 
&\;&\frac{1}{2}R\ddot R-\frac{1}{2}\dot R^2 
+\frac{1}{\rho_L}P_a 
+2\dot R\dot {\cal R}(1-\frac{\dot R}{c})+R\ddot {\cal
R}(1-4\frac{\dot R} {c})-\frac{2}{c}R\dot {\cal R}\ddot R-
\frac{1}{c}R^2\stackrel{...}{\cal R} \nonumber \\ [2ex] 
&\;&= 
h_G[P_G]+h_L[P_L] 
-2\frac{\eta_L}{\rho_L}\left\{ 
[\frac{1}{R}-\frac{1}{2c}\partial_t]v_a(r,t)
\right\}_{r=R}+{\rm O}\left ((\dot R/c)^2\right ) 
\;\;. \end{eqnarray} 
We observe that for $\dot {\cal R}\equiv 0$, i.e. when the bubble
wall velocity is identical to the incoming acoustical velocity
field at $r=R(t)$, all terms involving $1/c$-corrections vanish.
However, on the r.h.s. the viscous damping term survives, which
involves only the incoming sound field. This is to be expected,
since no outgoing sound wave is generated, in agreement with the
nonperturbative results for the velocity and pressure fields,
$v=v_a\equiv\partial_r\phi_{in}$ and $P=P_a\equiv -
\rho_L\partial_t\phi_{in}$ 
in this case. 
 
We arrive at a more useful form of the bubble equation of motion by
eliminating $\stackrel {...}{R}$ from Eq.\,(\ref{Reqperturb}). This
can be achieved at the same order in $\dot R/c$ by using the
equation at leading order (i.e. for $c\rightarrow\infty$) to
calculate $\ddot R$ and then $\stackrel {...}{R}$\,: 
\begin{equation} \label{dddR} 
R\stackrel{...}{R}=\frac{2}{3}\left\{
\frac{d}{{d}t}[{\rm r.h.s.}|_\infty ]-\frac{9}{2}\dot R\ddot R -
\frac{1}{\rho_L}\dot P_a+2\ddot Rv_a+3\dot R\dot v_a+R\ddot v_a
\right\} 
\;\;, \end{equation} 
where the term $[{\rm r.h.s.}|_\infty ]$ denotes the r.h.s. of
Eq.\,(\ref{Reqperturb}) in the limit $c\rightarrow\infty$ and
$P_a\equiv P_a(R,t)$\,, $v_a\equiv v_a(R,t)$\,, $\dot
P_a\equiv\frac{d}{{d}t}P_a(R,t)$\,, $\dot v_a\equiv
\frac{d}{{d}t}v_a(R,t)$\,, etc., from now on. Reinserting this
expression into Eq.\,(\ref{Reqperturb}), we finally obtain the
equation:
\begin{eqnarray} \label{Reqfinal} 
&\;& 
R\ddot R\left ((1+\underline{\frac{1}{2}})(1-2\frac{\dot R}{c})
+\underline{\frac{2}{3}}
\frac{v_a}{c}\right )
+\frac{3}{2}\dot R^2\left (1-\frac{4}{3}\frac{\dot R-v_a}{c}\right
) \nonumber \\ [2ex] 
&\;& 
-2\dot Rv_a-R\dot v_a\left (1-(1+\underline{1})\frac{\dot
R}{c}\right ) +\underline{\frac{1}{3}}\frac{R^2}{c}\ddot v_a 
\nonumber \\ [2ex] 
&\;&-
\left [1+(1-\underline{\frac{1}{3}})\frac{R}{c}\frac{d}{{d}t}
|_{c\rightarrow\infty} 
\right ]
\left [ 
h_G[P_G]+h_L[P_L] 
-2\frac{\eta_L}{\rho_L}\left\{ 
[\frac{1}{R}-\frac{1}{2c}\partial_t]v_a(r,t)
\right\}_{r=R}
-\frac{1}{\rho_L}P_a 
\right ] 
\nonumber \\ [2ex] 
&\;&  
+{\rm O}\left ((\dot R/c)^2\right )\;=\;0
\;\;, \end{eqnarray} 
where we separated out the underlined terms, in order to facilitate
the discussion of our result; the term $\propto
{d}/{d}t|_{c\rightarrow\infty}$ is to be read as: take this limit
of the following expression before evaluating the derivative.  
Equation (\ref{Reqfinal}) generalizes previously considered
variants of the Rayleigh-Plesset equation. In comparison to earlier
related work, particularly
Refs.\,\cite{KellerMiksis,Loefstedtetal,Sco97}, we draw attention
to the following:
\begin{itemize} 
\item 
The dynamics of the gas in the interior of the bubble as well as of
the exterior liquid is fully incorporated in our derivation. It is
based on the Navier-Stokes equation (\ref{NavierStokes})
supplemented by the boundary condition (\ref{stressboundary1})
relating the normal components of the stress tensor at the bubble
surface. Neglecting the bubble interior amounts to setting the
underlined numerical constants in Eq.\,(\ref{Reqfinal}) to zero,
equivalent to neglecting $R\ddot R/2$ in Eq.\,(\ref{Req}), and
setting $h_G[P_G]\equiv 0$ in both equations. These apparently are
sizeable corrections which were not considered previously. However,
these terms cancel exactly, if the homologous dynamics of Section
3 is applicable for the bubble interior. Also, according to
Eq.\,(\ref{enthinside}), $h_G=0$ for a homogeneous bubble.
\item 
The influence of the external acoustic driving field is
consistently taken into account in our treatment of the sound field
outside of the bubble. The terms $\propto v_a$, i.e. the incoming
velocity field, or its derivatives, were neglected before; only
terms involving the driving pressure $P_a$ had been obtained. These
additional terms are necessary to restore the boundary condition on
the sound field at $r=R(t)$, see the comment after
Eq.\,(\ref{Reqperturb}). 
\item 
A realistic EoS for the liquid, Eq.\,(\ref{EOSL}), has been
employed here. For $n\gg 1$ the enthalpy $h_L[P_L]$ in
Eqs.\,(\ref{Req}),\, (\ref{Reqfinal}) reduces to $(P_L(R)-
P_0)/\rho_L$, which was used earlier instead. -- The viscous term
$\propto\eta_L$ was not studied before. 
\item 
The pressure boundary condition (\ref{stressboundary1}) contains
terms which have been neglected before: 
\begin{eqnarray} \label{stressboundaryfinal} 
P_L(R)&=&P_G(R)-\frac{2\sigma}{R}-(3\eta_{bG}+4\eta_{sL})\frac{\dot
R}{R} \nonumber \\ [2ex] 
&\;& 
+(\frac{4}{3}\eta_{sL}+\eta_{bL})(\frac{2}{R}v_a-
\frac{1}{c}\{\partial_t v_a(r,t)\}_R)+{\rm O}\left ((\dot
R/c)^2\right ) 
\;\;, \end{eqnarray} 
where we used Eq.\,(\ref{dvelocitydr}) and inserted the {\it s}hear
and {\it b}ulk viscosities for liquid and gas explicitly. Except
for a highly compressed or, rather, very hot gas \cite{Pierce}, its
bulk viscosity presents a negligible contribution in
Eq.\,(\ref{stressboundaryfinal}), being on the \%-level as compared
to liquid water at most. However, we also note the corrections 
due to the incoming sound field. 
\end{itemize} 
If we discard altogether the modifications just discussed, we
reproduce Eqs.\,(9) and (10) of Ref.\,\cite{Loefstedtetal}. In
Section 5 we will explore numerically the effects of the
corrections presently discussed.

\subsection{The Driving Sound Field} 
Having further studies of single bubble sonoluminescence in mind,
we determine here those quantities entering our results which
depend on the external acoustic driving field. For typical
experiments on single bubble sonoluminescence carried out to date
the driving sound frequency is approximately $\nu\approx
25\,$Khz\,, which translates into a wavelength of $\lambda\approx
5.9\,$cm given the sound velocity in water of $c=1481\,$m/s\,.
Relevant bubble radii are such that $R/\lambda\stackrel{<}{\sim}
10^{-3}$ \cite{Loefstedtetal,lightscattering}. 
Therefore, we consider the external field in the long-wavelength
limit. 
 
Following Ref.\,\cite{Loefstedtetal}, we assume a plane standing
wave: $\phi =A\sin (kz)\cos (\omega t)$\,, i.e. along the $z$-axis
with $\omega\equiv 2\pi\nu$ and $k\equiv 2\pi /\lambda =\omega
/c$\,. In the long-wavelength limit, with $r\stackrel{<}{\sim} R\ll
1/k$\,, the spherical component of the external field, $\phi_{in}$, 
determines the dominant driving force for spherically symmetric
bubble oscillations. Expressing $z=z_0+r\cos\theta$, where $z_0$
denotes the presently irrelevant position of the center of the 
bubble, we can project out the spherically symmetric monopole 
component of the driving field to obtain: 
\begin{equation} \label{phimono} 
\phi_{in}(r,t)=\frac{A}{2}\cos (\omega t)
\int_{-1}^1{d}(\cos\theta )\;{\rm P}_0(\cos\theta ) 
\sin (k[z_0+r\cos\theta ])
=A\sin (kz_0)\cos (\omega t)\frac{\sin (kr)}{kr} 
\;\;, \end{equation} 
with ${\rm P}_0\equiv 1$ denoting the appropriate Legendre
polynomial. 
 
We recall that $P(R,t)$ contains the contribution of the incoming
sound field, i.e. the external ``acoustic driving pressure'':
\begin{eqnarray} \label{Pdrive} 
P_a(r,t)&\equiv&-\rho_L\partial_t\phi_{in}(r,t) 
\nonumber \\ [2ex] 
&=&P_A\sin (\omega t)\frac{\sin (kr)}{kr} 
=P_A\sin (\omega t)\left (1-\frac{1}{3!}(kr)^2+{\rm
O}((kr)^4)\right )  \;\;, \end{eqnarray} 
cf. Eq.\,(\ref{pressure}). The amplitude of the acoustical pressure
field is $P_A\equiv\omega\rho_LA\sin (kz_0)$\,. Similarly, the
velocity field $v(r,t)$ has an external contribution: 
\begin{eqnarray} \label{vdrive} 
v_a(r,t)&\equiv&\partial_r\phi_{in}(r,t) 
\nonumber \\ [2ex] 
&=&\frac{P_A}{c\rho_L}\cos (\omega t)\left (
\frac{\cos (kr)}{kr}- 
\frac{\sin (kr)}{(kr)^2}\right ) 
=-\frac{P_A}{3c\rho_L}\cos (\omega t)\left (kr+{\rm
O}((kr)^3)\right ) \;\;, \end{eqnarray} 
cf. Eq.\,(\ref{velocity}), which also enters our results for the
bubble equation of motion and pressure boundary condition. In
general, both quantities, $P_a$ and $v_a$\,, are determined by the
experimental set-up, in particular the external ultrasound sources
which drive the bubble oscillations. 
 
Finally, we rewrite the relevant spherically symmetric part of the
external velocity potential as a superposition of an outgoing and
an ingoing spherical wave. We obtain: 
\begin{equation} \label{phispherical} 
\phi_{in}(r,t)=\frac{cP_A}{2\omega^2\rho_L}\{
\sin (\omega [t+r/c])-\sin (\omega [t-r/c])\}/r 
\;\;, \end{equation} 
which will be useful in the following section. 

\subsection{A Nonperturbative Bubble Equation of Motion} 
We present in this section the bubble equation of motion resulting
when we do not use the expansion in $\dot R/c$ which was
employed in Section 4.1\,, see Ref.\,\cite{KellerMiksis} for
earlier related work.
 
Since the velocity potential $\phi$ obeys the spherical wave
equation (\ref{sphwave}) for the presently considered case of a
spherically symmetric gas bubble immersed in a liquid, we can
immediately write down its most general form (see Section 2.2): 
\begin{equation} \label{phioutin} 
\phi (r,t)=\frac{1}{r}\left (
f(t_{ret})+g(t_{adv})-g(t_{ret})\right ) 
\;\;, \end{equation} 
with arbitrary functions $f$ and $g$ and where: 
\begin{equation} \label{tretadv} 
t_{ret}\equiv t-r/c\;\;,\;\;\;
t_{adv}\equiv t+r/c 
\;\;. \end{equation} 
The terms involving $g$ represent the external field, an example of
which was considered in the previous section, see
Eq.\,(\ref{phispherical}) in particular. Generally, the relative
sign between the retarded and advanced contributions results from
the requirement that the external field be regular at $r=0$\,, i.e.
the position of the center of the bubble. The term involving $f$
presents the additional outgoing (sound) contribution to the
velocity potential which is generated by the moving bubble wall.
 
Then, our task is to determine (or eliminate) the functions $f$ and
$g$ such that $\phi$ satisfies the velocity boundary condition
(\ref{boundary}): \begin{equation} \label{boundarygen} 
\dot R=\partial_r\phi |_R=-\frac{1}{cR}\left (
c\phi +f'(t_{ret})-g'(t_{adv})-g'(t_{ret})\right )_R 
\;\;, \end{equation} 
where $f'(x)\equiv {d}f(x)/{d}x$ etc., and where 
Eqs.\,(\ref{phioutin}),\,(\ref{tretadv}) were used to obtain its
present form. Similarly, we rewrite the pressure boundary condition
(\ref{stressboundary1}): 
\begin{equation} \label{stressboundarygen} 
P_L(R)-\eta_L\frac{1}{r}\partial_r^2(r\phi)|_R=
P_G(R)-(4\eta_{sL}+3\eta_{bG})\frac{\dot R}{R}-\frac{2\sigma}{R}
\;\;. \end{equation} 
Here we made use of the velocity boundary condition, i.e. the first
of Eqs.\,(\ref{boundarygen}), and assumed the linear velocity
profile inside the bubble which was introduced in Section 3\,,
$v=\partial_r\phi=r\dot R/R$ for $r\leq R$\,, which determines the
gas viscosity term $\propto\eta_{bG}$. As we pointed out already
after Eq.\,(\ref{stressboundary1}), only the bulk viscosity of the
gas contributes for a linear (homologous) velocity profile. 
 
Furthermore, as discussed in Section 4.1\,, we employ the 
Navier-Stokes equation Eq.\,(\ref{NavierStokes}) as the underlying
dynamical equation of motion of the gas and liquid fluids.
Integrating over the radial coordinate $r$\,, we obtained the
nonperturbative Eq.\,(\ref{Req}) before, which we presently rewrite
as: 
\begin{equation} \label{Reqgen} 
-\left (\partial_t\phi +\frac{1}{2}(\partial_r\phi )^2\right )_R
={\cal I}+h[P_L]
-\frac{\eta_L}{\rho_L}\frac{1}{r}\partial_r^2(r\phi )|_R 
\;\;, \end{equation} 
where $h[P_L]$ denotes the liquid enthalpy evaluated in
Eq.\,(\ref{enthoutside}) and we collected terms pertaining to the
bubble interior into: 
\begin{equation} \label{I} 
{\cal I}\equiv -\frac{1}{2}R\ddot R+h_G[P_G] 
\;\;, \end{equation} 
with the gas enthalpy $h_G$ from Eq.\,(\ref{enthinside}); note that
$h_G=0$ for a homogeneous density inside the bubble. We recall once
again that the contribution of the bubble interior was evaluated
assuming the linear velocity profile. The discussion following
Eq.\,(\ref{Req}) furthermore implies that ${\cal I}\equiv 0$ for
the case of exactly homologous bubble interior dynamics (cf.
Section 3\,). Equations
(\ref{boundarygen}),\,(\ref{stressboundarygen}), and (\ref{Reqgen})
completely determine the bubble motion, i.e. $R(t)$\,, as we shall
demonstrate now. To begin with, we simplify the r.h.s. of
Eq.\,(\ref{Reqgen}) by using: 
\begin{equation} \label{enthvisc} 
h_L[P_L] 
-\frac{\eta_L}{\rho_L}\frac{1}{r}\partial_r^2(r\phi )|_R 
=h_L[P_L
-\eta_L\frac{1}{r}\partial_r^2(r\phi )|_R] 
-\frac{1}{n}  
\frac{\eta_L}{\rho_L}\frac{1}{r}\partial_r^2(r\phi )|_R 
\frac{P_L-P_0-
\eta_L\frac{1}{r}\partial_r^2(r\phi )|_R}{P_0+P_1}+\,\ldots \;\;,
\end{equation} 
which follows from Eq.\,(\ref{enthoutside}). Neglecting the small
cross term (and higher order corrections) between viscosity and 
compressibility of the liquid as before, we eliminate $P_L$ and
replace the argument of $h_L[\ldots ]$ with the help of the
pressure boundary condition (\ref{stressboundarygen}). The result
is: 
\begin{equation} \label{hL}  
h_L[
P_G(R)-(4\eta_{sL}+3\eta_{bG})(\dot R/R)-(2\sigma /R)] 
]\equiv h_L 
\;\;, \end{equation} 
which we henceforth abbreviate as indicated. 
 
In order to proceed, we calculate: 
\begin{eqnarray} \label{dtphigen} 
\partial_t\phi |_R&=&\frac{1}{R}\left (
f'(t_{ret})+g'(t_{adv})-g'(t_{ret})\right )_R
\nonumber \\ [2ex] 
&=&\frac{1}{R}\left (
-cR\dot R-c\phi +2g'(t_{adv})\right )_R 
\;\;, \end{eqnarray} 
where we used the boundary condition (\ref{boundarygen}) in the
second step. Inserting this result together with $\partial_r\phi
|_R =\dot R$ and the above evaluation of $h_L$ back into
Eq.\,(\ref{Reqgen}), we obtain: 
\begin{equation} \label{Reqgen1} 
c\dot R+\frac{c}{R}\phi |_R -\frac{2}{R}g'(t_{adv})|_R-
\frac{1}{2}\dot R^2 ={\cal I}+h_L 
\;\;. \end{equation} 
We have to eliminate $\phi$, in order to arrive at the closed
equation for $R(t)$\, which we are after. 
 
Taking the time derivative of Eq.\,(\ref{Reqgen1}), employing
Eq.\,(\ref{dtphigen}) once more, and solving the resulting equation
for $\phi$, one finds explicitly: 
\begin{equation} \label{phiexplicit} 
\frac{c}{R}\phi |_R = \frac{R}{1+\dot R/c}\left ( 
\ddot R(1-\frac{\dot R}{c})-\frac{c\dot R}{R}+\frac{\dot R^2}{R} -
\frac{1}{c}\frac{{d}}{{d}t}({\cal I}+h_L)\right )
+2\frac{g'}{R}-2\frac{g''}{c}
\;\;, \end{equation} 
where $g'$ stands for $g'(t_{adv})|_R$ and $g''$ similarly.
Reinserting this result into Eq.\,(\ref{Reqgen1}) yields:
\begin{equation} \label{Reqgen2}  
c\dot R+ 
\frac{R}{1+\dot R/c}\left ( 
\ddot R(1-\frac{\dot R}{c})-\frac{c\dot R}{R}(1-\frac{\dot R}{c}) -
\frac{1}{c}\frac{{d}}{{d}t}({\cal I}+h_L)\right ) -\frac{2}{c}g''-
\frac{1}{2}\dot R^2={\cal I}+h_L 
\;\;, \end{equation} 
from which equation we will derive our final results.
 
First of all, we obtain a perturbative equation of motion for
$R(t)$ by expanding the factor $(1+\dot R/c)^{-1}$ in powers of
$\dot R/c$ and collecting terms. Taking up to second order
corrections into account, the result is: 
\begin{eqnarray} \label{Reqgenperturb} 
&\;& 
R\ddot R\left (1-2\frac{\dot R}{c}+2(\frac{\dot R}{c})^2\right )
+\frac{3}{2}\dot R^2\left (1-\frac{4}{3}\frac{\dot
R}{c}+\frac{4}{3} (\frac{\dot R}{c})^2\right )-
\frac{2}{c}g''(t+R/c) 
\nonumber \\ [2ex] 
&\;&= 
\left (1+\frac{R}{c}(1-\frac{\dot R}{c})\frac{{d}}{{d}t}\right )
\left ({\cal I}+h_L\right )+{\rm O}\left ((\dot R/c)^3\right )
\;\;. \end{eqnarray} 
It is straightforward to calculate higher order corrections in this
case, if necessary. Equation (\ref{Reqgenperturb}), when evaluated
including only first order corrections in $\dot R/c$\,, essentially
agrees with our previous perturbative result,
Eq.\,(\ref{Reqfinal}). To see this, note that ${d}{\cal I}/{d}t$
generates a term $\propto\stackrel{\ldots}{R}$\,, using
Eq.\,(\ref{I}). The third order derivative can then be eliminated
similarly as between Eqs.\,(\ref{Reqperturb})--(\ref{Reqfinal}). 
 
However, a crucial difference remains in that presently we treat
the external field exactly, which is represented here by the term
$\propto g''$\,, whereas in Section 4.1 the external field was
treated perturbatively on the same footing as the sound field
generated by the bubble wall motion. By comparison of
Eqs.\,(\ref{phispherical}) and (\ref{phioutin}) (the external field
terms involving $g$\,) we find here explicitly: \begin{equation}
\label{ddg} 
-\frac{2}{c}g''(t+R/c)=\frac{P_A}{\rho_L}\sin (\omega [t+R/c])
\;\;, \end{equation} 
for the example of an external standing plane wave pressure field
discussed in Section 4.2\,. We remark that $\dot R/c$ may vary
systematically between less than about 1\,ns and about 100\,ns
during a nonlinear oscillation cycle with bubble parameters in the
SBSL regime. Neglecting this retardation in the external field
term, terms related to the bubble interior, and modifications due
to the realistic liquid EoS and enthalpy, Eqs.\,(\ref{EOSL}) and
(\ref{enthoutside}), which enter our derivations, we again
reproduce Eqs.\,(9) and (10) of Ref.\,\cite{Loefstedtetal}, if we
truncate Eq.\,(\ref{Reqgenperturb}) at O$(\dot R/c)$\,.
 
Alternatively, instead of expanding Eq.\,(\ref{Reqgen2}), we simply
multiply this equation by the denominator $1+\dot R/c$ to obtain:
\begin{equation} \label{Reqgen3} 
R\ddot R(1-\frac{\dot R}{c})
+\frac{3}{2}\dot R^2(1-\frac{1}{3}\frac{\dot R}{c})
-\frac{2}{c}(1+\frac{\dot R}{c})g''(t+R/c)
=(1+\frac{\dot R}{c})({\cal I}+h_L) 
+\frac{R}{c}\frac{{d}}{{d}t}({\cal I}+h_L) 
\;\;, \end{equation} 
which presents the nonperturbative bubble equation of motion; it
agrees with the result of Ref.\,\cite{KellerMiksis} in the limit of
the idealized adiabatic liquid EoS employed there 
and neglecting the bubble interior
(except for the pressure boundary condition).

\section{Numerical Results}
We now turn to a numerical study of characteristic properties of a
single air bubble in water at standard laboratory conditions (STP),
i.e. at the ambient temperature $T=293$\,K and under the ambient
pressure $P_0=1\,\mbox{atm}=1.01325\cdot 10^5\,\mbox{kg/m}\cdot
\mbox{s}^2$\,. Solutions of the nonperturbative equation of motion
(\ref{Reqgen3}) will serve as our standard, with which other
results shall be compared. 
 
The following material `constants' of water (STP) will be used
\cite{Pierce}: density $\rho_L=10^3\,\mbox{kg}/\mbox{m}^3$\,,
surface tension $\sigma =0.073\,\mbox{kg}/\mbox{s}^2$ (water/air),
sound velocity $c=1481\,\mbox{m}/\mbox{s}$\,. This latter value is
about 2\% higher than the one which follows from the adiabatic
water EoS (\ref{EOSL}) at STP with the parameters given in
Ref.\,\cite{Neppiras}, $n=7$ and $P_1=3\cdot
10^8\,\mbox{kg/m}\cdot\mbox{s}^2$\,, which we employ in
Eqs.\,(\ref{enthoutside}),\,(\ref{hL}). The shear and bulk
viscosities of water (STP) are \cite{Pierce}: $\eta_{sL}=1.002\cdot
10^{-3}\,\mbox{kg}/ \mbox{m}\cdot\mbox{s}$\,,
$\eta_{bL}=2.91\eta_{sL}$\,. 
 
The `standard' air bubble we study has an equilibrium radius
$R_0=4\cdot 10^{-6}$\,m\,. The adiabatic index is $\gamma =1.4$\,,
characteristic of di-atomic molecules. The equilibrium radius and
the particle number $N$ in the bubble, or equilibrium gas density
$\rho_0$, are related through the van\,der\,Waals EoS (\ref{EOSW})
and pressure boundary condition (\ref{stressboundarygen}) with
$P_L=P_0$ (STP), which corresponds to the following set of
parameters \cite{Pierce}: $N=0.913\cdot 10^{10}$\,, van\,der\,Waals
excluded volume $\frac{4}{3}\pi a^3=0.04\,\mbox{l}/\mbox{mole}$\,,
ratio of equilibrium to hard core density $\rho_0/\rho_a=0.0023$\,.
For the air viscosities we take \cite{Pierce}: $\eta_{bG}=
0.6\eta_{sG}=1.08\cdot 10^{-5}\,\mbox{kg}/\mbox{m}\cdot\mbox{s}$\,,
which indeed make a negligible contribution in the context of our
study.
 
The applied driving sound field, cf. Section 4.2, will be chosen to
oscillate with the frequency $\nu =\omega /2\pi =26\,$Khz in all
examples. Its amplitude will be fixed at $P_A=1.35P_0$, except when
stated otherwise. 
 
In the numerical calculations we consider the above specified
parameters as prescribed constants, but keep in mind that e.g. the
sound velocity and surface tension may change appreciably. The use
of a constant adiabatic index has been questioned before
\cite{Bar97}, since various transport effects leading to entropy
production may become important. We follow Ref.\,\cite{Bar97} in
that we artificially increase all viscosities by a factor three,
which has been found necessary to fit the standard Rayleigh-Plesset
equation, i.e. $R(t)$, to the experimental data obtained by
lightscattering methods \cite{lightscatteringnew,lightscattering}. 
 
In Fig.\,1 we show the full cycle of the time dependent radius
$R(t)$ of the air bubble in water (full line), the bubble surface
velocity $\dot R(t)$ (dashed line), and the sinusoidal external
driving pressure $P_a(r=0,t)$ according to Eq.\,(\ref{Pdrive})
(overlaid full line together with zero line, arbitrary units),
which coincides with the relevant long-wavelength limit. We
integrated numerically the nonperturbative Eq.\,(\ref{Reqgen3})
for a homologous bubble, in which case ${\cal I}=0$ according to
Eq.\,(\ref{I}). However, the gas pressure entering through
Eq.\,(\ref{hL}) is evaluated here for a homogeneous bubble
for the moment; we shall consider the effect of the
inhomogeneity in Section 5.3\,. We note that on the scale of
Fig.\,1\,, the $R(t)$ curves calculated with the other equations
derived in this work or the one used before in
Refs.\,\cite{Loefstedtetal,ChuLeung,Sco97,WuR93}, for example,
cannot be discriminated for our typical parameter set. Therefore,
we will in the following pay more attention to $\dot R(t)$\,, which
is much more sensitive to the differences between the various
equations of motion. 

\subsection{Perturbative vs. Nonperturbative Equations of Motion}
The equations of motion which have been used to study driven gas
bubbles in liquids in the SBSL parameter regime incorporated
effects of the sound emission only in the lowest order in $\dot
R/c$ so far, see e.g. Refs.\,\cite{Bar97,Loefstedtetal,Sco97} and
further references therein. In view of the high velocities of the
bubble surface reached during the collapse phase, which may reach
$\dot R/c\approx 1$ ($\equiv\mbox{Mach}\,1$), as indicated in
Fig.\,1\,, we study here how the various approximate treatments of
sound damping compare to the nonperturbative Eq.\,(\ref{Reqgen3}).
The bubble interior is treated here as explained in the context of
Fig.\,1 above. The individual sections in Fig.\,2 show the surface
velocity $\dot R(t)$ at the first bounce (left column) and the
second bounce (second column) for different amplitudes of the
driving pressure (top to bottom; $P_A$ in units of $P_0$); the
origin of the time-axis is arbitrarily chosen for the first bounce,
the delay until the corresponding one for the second bounce is
indicated in each case. The results are obtained from the
nonperturbative equation (full line), and its lowest order in 
$\dot R/c$ perturbative expansion (dashed line), cf.
Eq.\,(\ref{Reqgenperturb}), respectively. In the latter case also
the retardation shifting the time argument of the driving pressure
term, see Eq.\,(\ref{ddg}), has been neglected, in conformity with
earlier work. 
 
It is obvious from Fig.\,2 that characteristic phase shifts result
w.r.t. to the cycle timing set by the external driving pressure and
between the first and subsequent bounce(s). Physically most
important is the fact that within the perturbative treatment the
maximal surface velocity during the collapse is overestimated by a
considerable amount, e.g. $\approx 250\,\mbox{ms} ^{-1}$ for the
first bounce at $P_A=1.35P_0$\,. This implies that the collapse is
significantly less violent than has been concluded from previous
theoretical treatments of the bubble dynamics. In particular,
quantitative SBSL estimates derived from models which couple the
perturbatively corrected Rayleigh-Plesset equation, cf.
Eq.\,(\ref{Reqfinal}), to a full hydrodynamic simulation of the
bubble interior may need a revision \cite{ChuLeung,Chu,WuR93}. We
are led to this conclusion also from the various other corrections
to be discussed shortly. 
 
The origin of the retardation of the first bounce in the
perturbative approach is shown in Fig.\,3 to be due to the neglect
in the perturbative expansion of the time retardation effect. When
the retardation is included as in Eq.\,(\ref{ddg}), then the
positions of the first perturbative and nonperturbative bounce
coincide. However, the perturbative expansion including the
retardation effect continues to introduce an overestimate of the
collapse velocity. This is leading to a remaining phase shift of
15-20 ns between first and subsequent bounce(s), as seen in the
bottom portion of Fig.\,3\,.

\subsection{Homologous vs. Homogeneous Bubble Interior}
In order to demonstrate the influence of the behavior of the bubble
interior on the overall dynamics of the coupled bubble-liquid
system, beyond just providing the pressure resisting the collapse,
we consider here how the homologous bubble
description developed in Section 3 compares with the usual simpler
homogeneous matter distribution model of the bubble interior. 
 
We exploit here the presence of a fixed point $\xi_*$,
Eq.\,(\ref{rhostar}), where the density remains nearly constant.
Presence of this fixed point cannot be expected in general, 
however, we presently make this assumption based on our
numerical experience with the description of homologous and 
adiabatic changes of the interior distributions. We can evaluate
Eq.\,(\ref{Eulerint1}) conveniently integrating from $\xi_*R$ to
$R$.  With $P_*\equiv P(\rho _*)$, see Eq.\,(\ref{EOSW}), which
denotes the gas pressure of the corresponding homogeneous bubble,
we obtain the following approximate result for the homologous gas
pressure at the bubble surface: 
\begin{equation} \label{Phomolog}
P_G(R)=P_*\left (1-\frac{\gamma -1}{2}\frac{(1-\xi _*^2)\rho
_*R\ddot R} {(1-\rho _*/\rho _a)P_*}\right )^{\frac{\gamma}{\gamma
-1}} \;\;. \end{equation} 
  
Employing this expression for the gas pressure in the
nonperturbative equation of motion (\ref{Reqgen3}), we find the
results depicted in Fig.\,4 (full line), where also the behavior of
a homogeneous bubble (dashed line) is shown for comparison. We
observe only a rather small effect on $R(t)$, except for a 1.5\,ns
shift of the minimum. However, similarly to the effects illustrated
in Fig.\,2\,, we find here another sizeable reduction of the
maximal collapse speed ($\approx 200\,\mbox{ms} ^{-1}$). The
maximum of the gas pressure at the homologous bubble surface is
reduced by about 5\%\,, as compared to the homogeneous one, which
amounts to $\approx 10^3$\,atm\,. However, the maximal gas density
at the surface is much less affected and reaches approximately one
half liquid density in our calculational example.
 
\subsection{The Sound Field of a SBSL Bubble} 
As a further application of our study of the sound field imposed on
and rescattered by a typical SBSL bubble we evaluate it at a given
distance away from the bubble, where it can be measured by a
hydrophone, which preferably should be sensitive to the highest
expected sound frequencies \cite{Bar97,lightscatteringnew}. 
 
Perturbatively, it is most convenient to obtain the sound pressure
using Eq.\,(\ref{pressure}) or (\ref{pressureout}) in the form:
\begin{equation} \label{pressureoutpert}
 P(r,t)=P_a(r,t)+\frac{\tilde R}{r}\left (
P(\tilde R,\tilde t)-P_a(\tilde R,\tilde t)\right )
\;\;, \end{equation} 
where $\tilde R\equiv R(\tilde t)$ and $\tilde t$ can be further
evaluated with the help of Eq.\,(\ref{ttilde1}) or
Eq.\,(\ref{ttilde2}), whichever applies. The pressure difference on
the r.h.s. here can then be computed easily from the solution of
the perturbative equation of motion (\ref{Reqfinal}), cf. Section
5.1\,, since it is implicitly part of it.
 
The result at the lowest nontrivial order in $\dot R/c$ is shown in
Fig.\,5\,, for $r=1$\,mm away from the center of the bubble. We
clearly see the outgoing compression spikes riding on the
sinusoidal driving pressure, which are caused by the bubble
collapse and successive bounces. On the pressure scale of this
figure (cutting off the first maximum) the nonperturbative result
would be indistinguishable from the perturbative calculation. 
 
We calculate the pressure field nonperturbatively beginning with
Eq.\,(\ref{pressureoutpert}), as before. However, we proceed in
this case by using the pressure boundary condition,
Eq.\,(\ref{stressboundarygen}), in order to calculate $P(R,t)\equiv
P_L(R)$. The term $\propto -\eta _L$, in particular, can be
evaluated in the long-wavelength limit (neglecting a cross term
between viscosity and compressibility) to yield $+2\omega\eta
_LP_A\cos (\omega t)/(3c^2\rho _L)$\,, while all other terms are
straightforward to obtain from the numerical solution of the
nonperturbative Eq.\,(\ref{Reqgen3}); cf. the remarks explaining
the calculation of Fig.\,1 in Section 5.1\,. 
 
In Fig.\,6 we show the nonperturbative evaluation of the first
sound pressure spike at $r=1$\,mm\,. For comparison, also the
somewhat earlier and considerably stronger spike given by the
perturbative calculation, cf. Fig.\,5\,, is shown. If we correct
the calculated amplitude here for the geometric dispersion, then
the pressure at $r=0.6\,\mu$m reaches $4.5\cdot 10^4$\,atm at
maximum. Applying the generic damping factor of $10^{-4}$ for the
absorption of a (300\,Mhz) pulse travelling 1\,mm\,, we arrive
approximately at the pressure amplitude measured experimentally for
an SBSL bubble under similar conditions as assumed in our
calculation \cite{Bar97,lightscatteringnew}. However, another
interesting aspect here is the rise time of the sound signal, i.e.
from one-half to maximum amplitude, which takes only 40\,ps (decay
time 260\,ps). This is about two orders of magnitude below what has
been resolved in the above cited experiments.  
   
Finally, in view of the pressure spike results presented here, it
seems worth while to check one of our basic assumptions, which is
the linear acoustic approximation, Section 2.1. It is underlying
the linear wave equation for the velocity potential,
Eq.\,(\ref{sphwave}), which forms the basis of our study of the
emitted sound field. Whereas dispersion and absorption can be
incorporated in a linearized approximation, genuinely nonlinear
effects are beyond our present scope \cite{Pierce}. The linear
approximation requires that perturbations of the ambient state of
the fluid are relatively small. Employing the adiabatic liquid EoS
(\ref{EOSL}) for water and considering the maximum of the pressure
amplitude obtained here, see Fig.\,6\,, we obtain a maximal
compression and ratio of sound speeds: 
\begin{equation} \label{ratios}
\mbox{max}\,\frac{\rho}{\rho _L}\approx 1.4\;\;,\;\;\;
\mbox{max}\,\frac{c(\rho )}{c(\rho _L)}\approx 2.7
\;\;, \end{equation}
respectively. Whereas the compression may appear to be quite
tolerable, the temporary increase of the sound speed and its
implications for the damping of the bubble motion, when the sound
pulse is launched from the collapsing bubble surface, obviously
deserve further study. 
  
Furthermore, one may wonder about the behavior of the water next to
the bubble surface, when it is exposed to the ``cold shock''
indicated by our results, i.e. several $10^4$\,atm pressure
increase within less than 50\,ps\,. 
 
\section{Summary, Conclusions and Outlook}
The aim of our present work has been to reconsider the sound
emission from the highly nonlinear, large amplitude motion of the
interface between the gas inside and the liquid outside a
cavitating bubble. Previous studies typically evaluated the
radiated sound field in lowest order of a perturbation expansion in
$\dot R/c$, i.e. valid for slow bubble wall motion as compared to
the sound velocity in the liquid. Generally, it is believed that
$\dot R(t)$ reaches up to (or even exceeds) the sound velocity for
externally driven bubbles in the parameter regime where
sonoluminescence is observed experimentally \cite{Bar97}. This
necessitates a nonperturbative treatment, such as ours, cf. Section
2.2 and see Eq.\,(\ref{pressure}), in particular. We anticipate
that the sound signal from a nonlinearly oscillating bubble wall
may provide an important additional diagnostic tool. It should
reflect the essential short-time scale(s) of this motion, which
vary over several orders of magnitude \cite{Loefstedtetal,Sco97},
with current technology limiting the resolution in the $(100\,{\rm
Mhz})^{-1}$ range, i.e. less than about 10\,ns\,. This may be
particularly valuable in cases where light scattering methods
\cite{lightscatteringnew,lightscattering} do not truly reflect the
motion of a sharply defined bubble wall or are not applicable at
all, such as for liquid metals \cite{Young}.
 
We recall that based on Section 3 we describe the bubble interior
using homologous profile functions for the density and velocity
distributions. These allow for a more realistic study of the
dynamics and particular properties of the high compression phase of
strongly driven bubble oscillations, the importance of which has
been shown earlier \cite{ChuLeung,Chu}. Presently we employ only a
simplified version of the semianalytic variational approach
developed in Refs.\,\cite{Sco97,Tak97}. In any case, our derivation
of Eq.\,(\ref{Reqgen3}) allows to incorporate easily any more
precise model of the bubble interior; and it may be possible to
extend this approach in order to account for the important effects
of heat conductivity and mass diffusion.
 
Our numerical examples obtained in Section 5 employ as another
approximation the homologous solutions with the usual adiabatic
van\,der\,Waals EoS. These EoS become questionable when the energy
density reaches the ionization regime. The dynamic behavior of the
gas mixtures within a rapidly oscillating bubble, here assumed to
remain that of a di-atomic gas (air), deserves further study. In
parallel to the cyclic in- and outgassing from the liquid into the
bubble (``rectified diffusion''), it remains to be seen whether an
essential amount of liquid vaporizes at the bubble surface and
recondenses during an oscillation cycle. Related effects may help
to explain the observed sensitivity of sonoluminescence to
experimental parameters such as temperature in particular 
\cite{Bar97,HillerWenningeretal,BarberWuetal}. In general,
transport phenomena within the bubble and across the phase boundary
have to be incorporated whenever the motion becomes fast compared
to characteristic relaxation times $\tau_i$\,, $\dot
R\approx\triangle_i/\tau_i$\,, where $\triangle_i$ denotes the
scale of a corresponding gradient (e.g. mass density, partial
pressure, temperature, etc.). In particular, shock waves may be
launched into the bubble interior \cite{Loefstedtetal,WuR93,Mos94}
or exterior \cite{ChuLeung}, the description of which is beyond the
scope of this work. 
 
One may speculate whether or not the liquid (water) may be trapped
in a metastable state w.r.t. solidification into a high-density
phase (of ice) and in which form the corresponding binding energy
would be released most efficiently. Presumably it stays at or close
to the ambient temperature \cite{ChuLeung}. Assuming an effective
sound velocity of 2000\,m/s\,, the 300\,ps mean half width of the
pressure pulse obtained in Section 5.3 corresponds to a spatial
shell width of 0.6\,$\mu$m\,, i.e. about the van\,der\,Waals hard
core radius employed in our calculations. Near the collapsed state
with $R\approx 0.6\,\mu$m such a shell contains about $10^{11}$
water molecules, i.e. only about an order of magnitude more
particles than assumed (in the gas) inside. Furthermore, the sound
pulse amplitude decreases exponentially away from the bubble
surface (besides geometric dispersion $\propto 1/r$) due to
absorption in the liquid \cite{Pierce}. Then, assuming an extremely
conservative decay $\propto\exp (-\Delta r\,[\mbox{mm}]/.434)$\,,
one immediately estimates that the energy dissipated in this first
shell (proportional to square of amplitude decrease) amounts to
about 3/1000 of the pulse energy. This is more than an order of
magnitude times the energy emitted in the form of visible light by
a SBSL bubble \cite{Bar97}. The small spatio-temporal extension of
the region where this energy has to be dissipated, and how,
definitely deserves further study. This should provide an improved
starting point for the exploration of essential aspects of the
violent bubble collapse which may help to elucidate the nature of
sonoluminescence. We hope to come back to one or the other of these
fascinating aspects of SBSL bubbles which are especially related to
their sound field in our future work.   
 
We note that while in principle the effect of a dynamic treatment
of the bubble interior on the behavior of $R(t)$\,, 
which we introduced in Section 5.2 and Fig.\,4,  should be part of a full hydrodynamic
simulation \cite{ChuLeung,Chu,WuR93}, the sound emission discussed
in Section 5.1 had not been treated properly in any of the
approaches prior to our work. In view of the numerical results
presented in Section 5 it is obvious that the nonperturbative
treatment of the emitted sound field as proposed here, which
accounts for about 50\% of the damping of the driven bubble
oscillations, is mandatory. In particular, the standard
perturbative calculations tend to overestimate the maximally
reached surface velocity during the collapse phase considerably.
Thus a crucial aspect of the ``preparation phase'' for the unknown
light emission process can be described more accurately employing
the nonperturbative equation of motion derived in Section 4.3\,,
Eq.\,(\ref{Reqgen3}). This seems particularly relevant for detailed
hydrodynamic studies of the bubble interior, in which the exterior
has been described by the perturbative Rayleigh-Plesset equation
before \cite{Loefstedtetal,ChuLeung,Chu,WuR93}. Our numerical
results for the nonperturbative sound pulse emitted by an air
bubble in water (Section 5.3) fit qualitatively the first
experimental observations reported in
Refs.\,\cite{Bar97,lightscatteringnew}. On the other hand, as we
pointed out at the end of Section 5.3, they also indicate the
limitation of our present derivations. Whereas the bubble equation
of motion, Eq.\,(\ref{Reqgen3}), is derived from the intrinsically
nonlinear Navier-Stokes equation, in the form of
Eq.\,(\ref{Reqgen}) of Section 4.3\,, our considerations of the
emitted sound field are based on the acoustic approximation (see
Section 2.2). Therefore, we still neglect important sound
absorption and dispersion effects in the liquid. Using the
realistic EoS for water, Eq.\,(\ref{EOSL}) in Section 4.1\,, we
indicated in Eqs.\,(\ref{ratios}) the compression of the liquid and
the corresponding increase in the density dependent sound velocity
which are induced by the outgoing pressure spike next to the bubble
surface. It seems desirable to study in the future the importance
of nonlinear effects on the propagation and fate of the extremely
strong sound pulse emitted.  
 
There are published results on full hydrodynamic simulations of the
coupled bubble-liquid system, which include especially the exterior
fluid \cite{Mos94}. However, to the best of our knowledge, so far
no attention has been paid to the unusual behavior of the liquid
which may be caused by the pressure spike of $10^4\dots 10^5\,$atm
launched from the bubble surface, independently of whether there
are shock waves generated in the bubble interior or not. We
obtained such amplitudes in typical examples with generic SBSL
parameters, where particularly their rise time of only about 40\,ps
seems quite astounding, but also reminiscent of the shortness of
the SBSL light pulses. 
 
\subsection*{Acknowledgement}  
We thank C.E. Aguiar, B. P. Barber, L. A. Crum, R. Donangelo, Y.
Hama, L.M. Pimentel, I. Scott and M. VanZeeland for stimulating
discussions. 
This research was supported in part by US-Department of Energy
under Grant No. DE-FG03-95ER40937, by NSF under grant INT-9602920,
by Brazil-PRONEX-41.96.0886.00 and by FAPERJ-Rio de Janeiro. One of
us (HTE) would also like to thank ITP-Santa Barbara for
hospitality, where this work was in part supported by NSF under
grant PHY94-07194.
 

 
\newpage
\subsection*{Figure Captions}
\vskip .5cm
\noindent 
Fig.\,1\,: Radius $R$ (full line) and surface velocity 
$\dot R$ (dashed line) as a function of time for an 
air bubble in water computed according to the nonperturbative
Eq.\,(\ref{Reqgen3}). The time dependent driving 
pressure with $P_A=1.35P_0$ is shown (overlaid together 
with zero line, arbitrary units). See main text for 
specification of other system 
parameters and further details. 
\vskip .5cm
\noindent
Fig.\,2\,: Surface velocity $\dot R(t)$ for the bubble of Fig.\,1
at first bounce (left column) and second bounce (second column) for
different amplitudes of driving pressure (top to bottom; $P_A$ in
units of $P_0$); the delay between origins of the time-axes between
the bounces is indicated. Results from the nonperturbative
Eq.\,(\ref{Reqgen3}) (full line) and the perturbative 
Eq.\,(\ref{Reqgenperturb}) (dashed line); see main 
text for further details.
\vskip .5cm
\noindent
Fig.\,3\,: Same as Fig.\,2\,, however, employing the
nonperturbative Eq.\,(\ref{Reqgen3}) (full line) and perturbative
Eq.\,(\ref{Reqfinal}), comprising time retardation (dashed line). 
\vskip .5cm
\noindent 
Fig.\,4\,: Comparison of the radius $R$ (top) and surface velocity
$\dot R$ (bottom) for the bubble of Fig.\,1\,, using
Eq.\,(\ref{Reqgen3}), for a homologous (full line) and a
homogeneous (dashed line) bubble. \vskip .5cm 
\noindent
Fig.\,5\,: The pressure amplitude at $r=1$mm from the center of the
bubble of Fig.\,1\,, calculation including O($\dot R/c$) corrections; see
main text for further details.
\vskip .5cm 
\noindent 
Fig.\,6\,: The first spike of the outgoing compression pulse
launched by the collapsing bubble, see Fig.\,5\,. For comparison
the result of the nonperturbative calculation (smaller, later
pulse) is shown together with the perturbative one at O($\dot
R/c$)\,. 
 
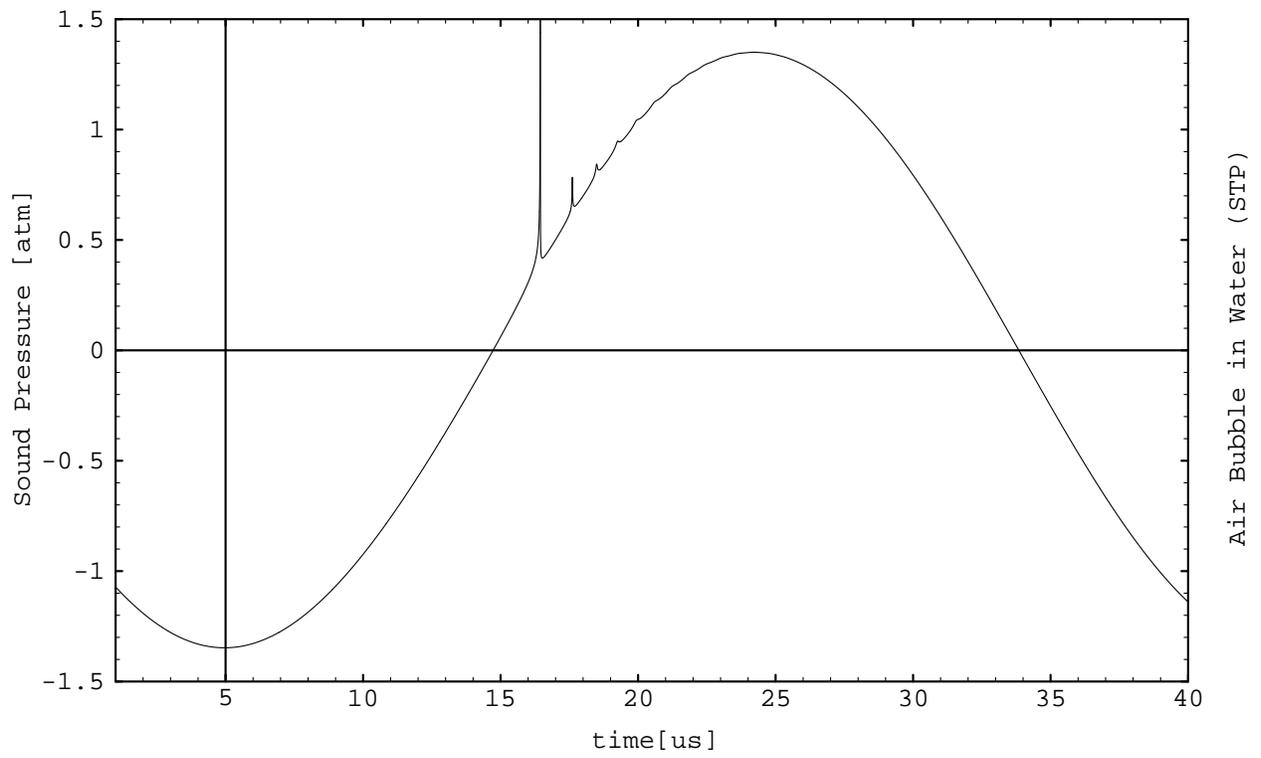
\end{document}